\documentclass{nature}
\usepackage[utf8]{inputenc}
\usepackage{graphicx}
\usepackage{graphics}
\graphicspath{ {./Figures/} }
\usepackage{float}
\usepackage{lineno}
\usepackage{comment}
\usepackage{tabularx}
\usepackage{booktabs}
\usepackage{multirow}
\usepackage{amsmath,amssymb}
\usepackage{bbold}
\usepackage[dvipsnames]{xcolor}
\usepackage{mathtools}

\usepackage{textcomp,gensymb}
\usepackage{prettyref}
\newrefformat{fig}{Fig.~\ref{#1}}
\newrefformat{exfig}{SI Fig.~\ref{#1}}

\usepackage{caption}
\DeclareCaptionLabelFormat{vertline}{#1 #2 $|$}
\makeatletter \let\saved@includegraphics\includegraphics
\AtBeginDocument{\let\includegraphics\saved@includegraphics} %
\renewenvironment{figure}{\@float{figure}}{\end@float}
\makeatother

\definecolor{orange}{rgb}{1,0.5,0}

\newcommand{\beginsupplement}{%
        \setcounter{table}{0}
        \setcounter{figure}{0}
        \renewcommand{\figurename}{SI Figure}
        \renewcommand{\tablename}{SI Table}
     }

\newcommand{\cCom}{\mathbin{\raisebox{0.5ex}{,}}}
\newcommand{\vk}{{\boldsymbol{k}}}
\newcommand{\vK}{{\boldsymbol{K}}}
\newcommand{\vq}{{\boldsymbol{q}}}
\newcommand{\vr}{{\boldsymbol{r}}}
\renewcommand{\vec}[1]{{\boldsymbol #1}} %
\newcommand{\nt}{\notag\\}

\DeclareMathAlphabet{\mathcald}{U}{dutchcal}{m}{n}
\SetMathAlphabet{\mathcald}{bold}{U}{dutchcal}{b}{n}
\DeclareMathAlphabet{\mathalt}{U}{dutchcal}{b}{n}

\title{Ascendance of Superconductivity in Magic-Angle Graphene Multilayers}
\author{Yiran Zhang$^{1,2,3*}$, Robert Polski$^{1,2*}$, Cyprian Lewandowski$^{2,3}$, 
Alex Thomson$^{2,3,4}$, Yang Peng$^{5}$, Youngjoon Choi$^{1,2,3}$, 
Hyunjin Kim$^{1,2,3}$, Kenji Watanabe$^6$, Takashi Taniguchi$^6$, 
Jason Alicea$^{2,3}$, Felix von Oppen$^7$, Gil Refael$^{2,3}$, 
and Stevan Nadj-Perge$^{1,2\dagger}$}
\begin{document}

\maketitle

\begin{affiliations}

  \item T. J. Watson Laboratory of Applied Physics, California Institute of
  Technology, 1200 East California Boulevard, Pasadena, California 91125, USA
  \item Institute for Quantum Information and Matter, California Institute of
  Technology, Pasadena, California 91125, USA
  \item Department of Physics, California Institute of Technology, Pasadena,
  California 91125, USA
  \item Department of Physics, University of California, Davis, California 95616, USA
  \item Department of Physics and Astronomy, California State University, Northridge, California 91330, USA
  \item National Institute for Materials Science, Namiki 1-1, Tsukuba, Ibaraki
  305 0044, Japan
  \item Dahlem Center for Complex Quantum Systems and Fachbereich Physik, Freie Universit\"at Berlin, 14195 Berlin, Germany
  \item[*] These authors contributed equally to this work
  \item[$^\dagger$] Correspondence: s.nadj-perge@caltech.edu
  
\end{affiliations}
\begin{abstract}

Graphene moir\'e superlattices have emerged as a platform hosting an
  abundance of correlated insulating, topological, and
  superconducting phases. While the origins of strong correlations and
  non-trivial topology are shown to be directly linked to flat moir\'e
  bands\cite{caoCorrelatedInsulatorBehaviour2018,
    serlinIntrinsicQuantizedAnomalous2019, zondinerCascadePhaseTransitions2020,
    choiElectronicCorrelationsTwisted2019, wongCascadeElectronicTransitions2020,
    choiCorrelationdrivenTopologicalPhases2021,
    nuckollsStronglyCorrelatedChern2020}, the nature and mechanism of superconductivity remain enigmatic.  
    In particular, only alternating twisted
  stacking geometries\cite{khalafMagicAngleHierarchy2019} of bilayer and
  trilayer graphene are found to exhibit robust superconductivity manifesting as
  zero resistance and Fraunhofer interference
  patterns\cite{caoUnconventionalSuperconductivityMagicangle2018,
    parkTunableStronglyCoupled2021, haoElectricFieldTunable2021}. Here we
  demonstrate that magic-angle twisted tri-, quadri-, and pentalayers 
  placed on monolayer tungsten diselenide exhibit flavour polarization and
  superconductivity. We also observe insulating states in the
  trilayer and quadrilayer arising at finite electric displacement fields,
  despite the presence of dispersive bands introduced by additional graphene
  layers. Moreover, the three multilayer geometries allow us to identify universal 
  features in the family of graphene moir\'e structures arising from the intricate 
  relations between superconducting states,  symmetry-breaking transitions, 
  and van Hove singularities. Remarkably, as the number of layers increases, 
  superconductivity emerges over a dramatically enhanced filling-factor range.
  In particular, in twisted pentalayers, superconductivity extends well beyond the filling of four 
  electrons per moir\' e unit cell, demonstrating the non-trivial role of the 
  additional bands. Our results highlight the importance of the interplay between 
  flat and dispersive bands in extending superconducting regions in graphene moir\'e 
  superlattices and open new frontiers for developing graphene-based superconductors. 
\end{abstract}

While a rich phase diagram of quantum electronic phases has been realized in
many graphene superlattice structures, robust superconductivity is so far
exclusive to twisted bilayer graphene (TBG)\cite{caoCorrelatedInsulatorBehaviour2018,
  caoUnconventionalSuperconductivityMagicangle2018} and twisted trilayer
graphene (TTG)\cite{parkTunableStronglyCoupled2021,haoElectricFieldTunable2021}. Striking
differences between TBG and TTG (e.g., Pauli limit violation\cite{caoPaulilimitViolationReentrant2021} and Bose-Einstein condensate type 
superconductivity\cite{parkTunableStronglyCoupled2021,haoElectricFieldTunable2021} observed in TTG) 
may serve as clues to the origin of their phenomenology; nevertheless, our ability 
to identify the truly universal features of these systems is ultimately limited 
by the relative dearth of robust superconducting moir\' e materials, suggesting that 
further progress lies not only in a better understanding of TBG and TTG, but also in the 
discovery of new superconducting systems.

\begin{figure}[p]
    \centering
    \includegraphics[width=14cm]{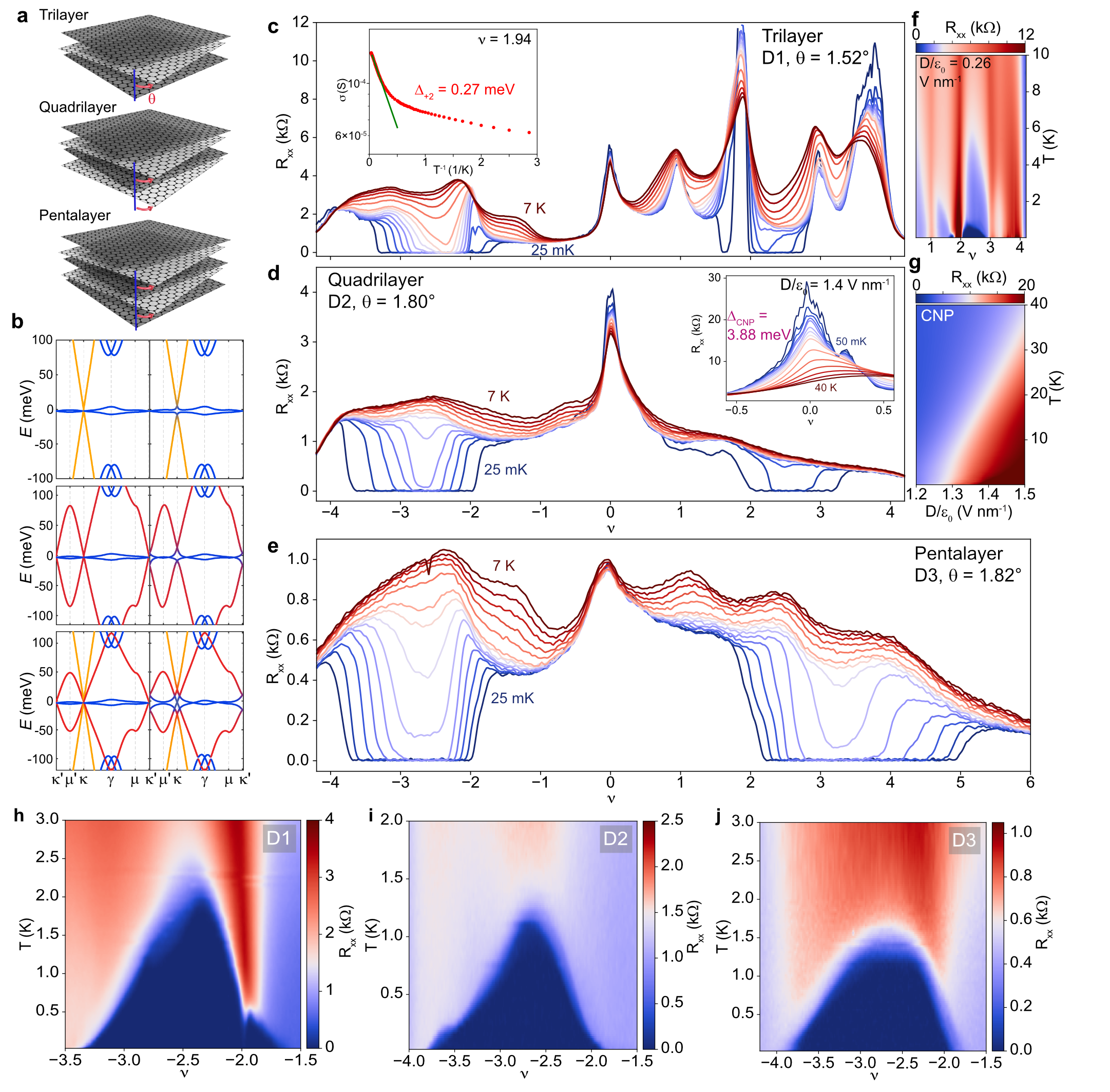}
    \caption{{\bf Superconductivity and correlated insulators in alternating twisted graphene
        multilayers.} {\bf a}, Schematics of the alternating twisted graphene
      multilayers where each successive layer is twisted by an angle $\pm\theta$
      relative to the previous one in an alternating sequence. {\bf b}, Band
      structure of twisted trilayer, quadrilayer, and pentalayer graphene (from
      top to bottom) for angles close to theoretical magic angle at zero $D$ field (left)
      and $D/\epsilon_0 \approx 0.4~\text{V nm}^{-1}$ (right) for valley K (see
      SI, section~\ref{methods:calculations}). {\bf c}--{\bf e}, Line cuts
      of $R_{xx}$ versus filling factor $\nu$ for a range of temperatures 
      (shown are traces taken first at $25$~mK, then every $0.25$~K from $0.25$~K to $2$~K, followed by every $1$~K from $3$~K to $7$~K),
      from top to bottom measured at
      $D/\epsilon_0 = 0.22~\text{V nm}^{-1}$ ({\bf c}), $-0.15~\text{V nm}^{-1}$
      ({\bf d}), and $0~\text{V nm}^{-1}$ ({\bf e}), respectively. Activation 
      gap fit of $\nu = +2$ TTG correlated insulator for 
      $D/\epsilon_0 = 0.26~\text{V nm}^{-1}$ is shown in inset of {\bf c}. 
      The inset of {\bf d} shows insulators in TQG at charge neutrality and larger electric fields. 
      {\bf f}, $R_{xx}$ versus temperature and $\nu$ for the trilayer 
      focusing around $\nu = +2$ at $D/\epsilon_0 = 0.26~\text{V nm}^{-1}$. {\bf g},
      $R_{xx}$ versus temperature and $D$ field for the quadrilayer focusing
      near charge neutrality. {\bf h}--{\bf j}, $R_{xx}$ versus temperature and $\nu$
      for hole doping, showing superconducting domes around $\nu = -2$ in the same systems and for the
      same $D$ fields as in {\bf c}--{\bf e}.}
    \label{fig:Fig1}
\end{figure}

We investigate twisted graphene multilayers where each successive layer is
twisted by an angle $\pm\theta$ relative to the previous one in an alternating
sequence (\prettyref{fig:Fig1}a). For an even number $n$ of layers, the spectrum
at zero displacement field $D$ is expected to separate into $n/2$
independent TBG-like bands, each characterized by a different effective twist
angle. When the number of layers $n$ is odd, in addition to $(n-1)/2$ TBG-like
bands, one monolayer-graphene-like (MLG-like) band (essentially a Dirac cone) is
expected\cite{khalafMagicAngleHierarchy2019} (see left column of \prettyref{fig:Fig1}b for examples when $n$ is 3, 4 and 5). The system may be conveniently
modified through the application of a displacement field $D$, which 
controllably hybridizes the different bands  (\prettyref{fig:Fig1}b right column).
Experimentally, we explore properties of alternating twisted trilayer,
quadrilayer, and pentalayer graphene (TTG, TQG, TPG) structures with
$\theta = 1.52\degree$ ~(device D1, trilayer), $\theta = 1.80\degree$ ~(D2,
quadrilayer), and $\theta = 1.82\degree$ ~(D3, pentalayer), respectively (see
Methods and Supplementary Information (SI), section \ref{methods:uniformity} 
for fabrication and twist-angle characterization). These twist angles all lie close
to the theoretically predicted ``magic'' values needed to obtain one set of flat
TBG-like bands
($\theta_{\rm TTG}^\mathrm{magic} = \sqrt{2}\theta_{\rm TBG}^\mathrm{magic}\approx 1.53\degree$,
$\theta_{\rm TQG}^\mathrm{magic}= (\sqrt{5}+1)\theta_{\rm TBG}^\mathrm{magic}/2\approx 1.75\degree$,
and
$\theta_{\rm TPG}^\mathrm{magic} = \sqrt{3}\theta_{\rm TBG}^\mathrm{magic}\approx 1.87\degree$
assuming an effective TBG twist angle
$\theta_\mathrm{TBG}^\mathrm{magic}=1.08\degree$; see SI, section~\ref{methods: continuum_model})\cite{khalafMagicAngleHierarchy2019}.
We find that TTG, TQG, and TPG all exhibit hallmark signatures
of strong correlations (\prettyref{fig:Fig1}c-e), including robust
superconductivity and flavour symmetry breaking as revealed by pronounced
resistance peaks around certain integer filling factors $\nu$ (number of electrons 
per moir\'e site; see SI, 
section \ref{methods:uniformity} for assignment of $\nu$).

In addition to the symmetry-breaking transitions previously reported in
TTG\cite{parkTunableStronglyCoupled2021, haoElectricFieldTunable2021, 
liuCoulombScreeningThermodynamic2021}, our TTG
structure (coupled to a coupled to tungsten diselenide (WSe$_2$) monolayer\cite{aroraSuperconductivityMetallicTwisted2020}) exhibits a previously unobserved
correlated insulating state near $\nu=+2$ at finite $D$ (inset in
\prettyref{fig:Fig1}c; see also 
\prettyref{exfig:CI_TTG} for more complete $D$ and $\nu$ dependence). 
This insulating state cannot arise at the non-interacting band theory level 
(\prettyref{fig:Fig1}b right column; also see SI, sections \ref{methods:ttg_tqg_gap} and
\ref{methods:calculations} for more data and further discussion) and is instead 
attributed to the interplay between an interaction-driven cascade transition 
and hybridization induced by the $D$ field (e.g., as captured by 
Ref.~\citenum{christosCorrelatedInsulatorsSemimetals2021, xieTwistedSymmetricTrilayer2021}). 
We have also detected an insulating state developing at finite $D$ fields in TQG 
near charge neutrality (\prettyref{fig:Fig1}d inset and \prettyref{fig:Fig1}g).
However, in contrast to TTG, the TQG insulating state can be explained through
the $D$-induced hybridization only. Importantly, the detection of insulating
gaps in TTG and TQG implies a low level of disorder in our samples 
(see also \prettyref{exfig:uniformity}).

The superconducting regions in all three structures extend over significantly
larger filling factor ranges in comparison to
TBG where superconductivity is typically observed within $2<|\nu|<3$. 
Moreover, as the layer number is increased, superconductivity on both
electron and hole sides persists to progressively higher fillings, reaching
$\nu\approx+5$ on the electron side for TPG (\prettyref{fig:Fig1}c-e).
Along with a zero longitudinal resistance $R_{xx}$ observed in the characteristic
$\nu$ vs. $T$ dome 
(\prettyref{fig:Fig1}h-j), we also measure well-resolved Fraunhofer-like 
patterns exhibiting large critical currents ($\sim400$~nA), substantiating 
the robustness of phase coherence (see \prettyref{exfig:Fraunhofer}). 
Moreover, high critical perpendicular magnetic fields $B_{c}$ (typically $\sim0.8$~T) 
indicate that the corresponding Ginzburg--Landau coherence lengths $\xi_\mathrm{GL}$
(approximately $10-30$~nm) are significantly smaller than those observed in 
TBG and deviate from the weak-coupling prediction,
$\xi_\mathrm{GL}\approx \hbar v_F/\pi \Delta$ with
$\Delta\approx1.76k_B T_c$---suggesting a strong-coupling origin of
superconductivity\cite{parkTunableStronglyCoupled2021,haoElectricFieldTunable2021}
(see SI, section \ref{methods:T_B_dependence}). When combined with other
recent experiments\cite{caoPaulilimitViolationReentrant2021,
  ohEvidenceUnconventionalSuperconductivity2021,
  kimSpectroscopicSignaturesStrong2021}, these observations affirm the
unconventional nature of superconductivity within the entire class of graphene
moir\'e systems. Further, the measurements on three to five layers indicate that
the addition of layers promotes superconductivity over a broader filling window
despite the coexisting dispersive bands as well as the ostensibly increased
vulnerability to disorder---both from the additional twist angles as well as
from the sensitivity to the relative displacement between layers.

\begin{figure}[p]
    \centering
    \includegraphics[width=16cm]{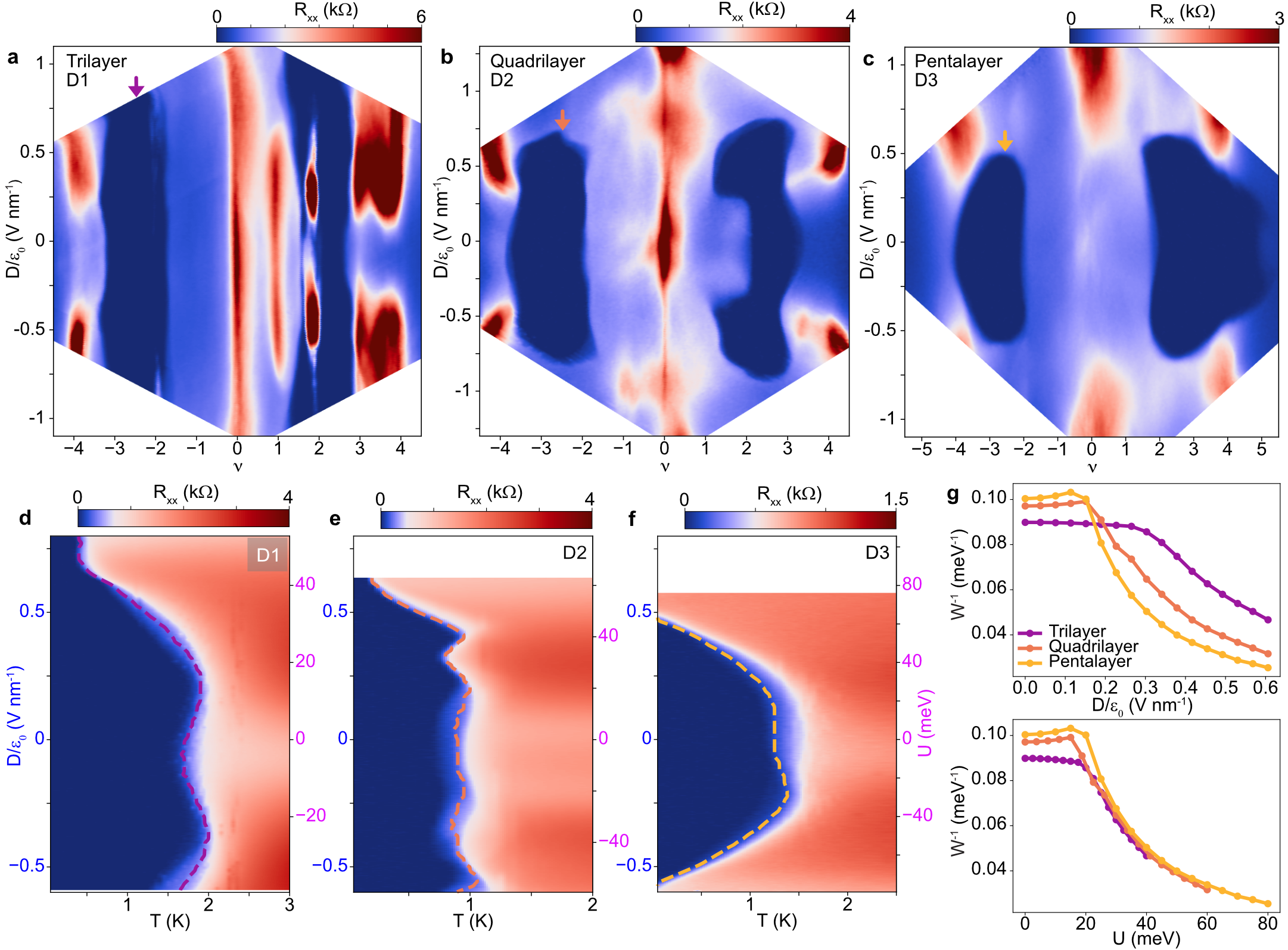}
    \caption{{\bf TTG, TQG and TPG phase diagrams and electric field-tunable superconductivity.}
      {\bf a}--{\bf c}, $R_{xx}$ versus filling factor $\nu$ and displacement
      field $D$ for twisted trilayer ({\bf a}), quadrilayer ({\bf b}), and
      pentalayer ({\bf c}) graphene, respectively. All data are taken at $25$~mK,
      and the dark blue regions signal superconductivity. 
      For electron-doped TTG and TQG, superconducting regions extend towards $\nu = +1$ at intermediate $D$ field. 
      {\bf d}--{\bf f}, $R_{xx}$ versus temperature and $D$ 
      (or equivalent potential difference $U$ between layers, see SI, 
      section \ref{methods:ttg_tqg_gap} for conversion from $D$ to $U$) for the filling 
      factors indicated by arrows in {\bf a}--{\bf c}. Critical temperature 
      $T_{c}$ is indicated by a dashed line that delineates $10\%$ of the normal 
      state resistance (see also SI, section~\ref{methods:T_B_dependence}). 
      $T_{c}$ is maximized at finite $D$ fields. 
      Overall, superconductivity is suppressed more easily with $D$ 
      as the layer number is increased for 
      both hole ({\bf d}--{\bf f}) and electron 
      (\prettyref{exfig:Tc_vs_D_electron}) doping. {\bf g}, Theoretical calculations of
      the inverse of the flat-band bandwidth for twisted
      trilayer, quadrilayer, and pentalayer graphene as a function of
      $D/\epsilon_0$ (top) and potential difference $U$ (bottom). 
      For a fixed $D$, the bandwidth of the flat bands is larger
      for systems with more layers, but when expressed as a function of $U$, 
      the flat-band broadening follows a
      similar trend across the different structures. See SI, section~\ref{methods: continuum_model}.}
    \label{fig:Fig2}
\end{figure}

In addition to the pronounced $\nu$-dependence, the observed superconducting
pockets are highly tunable with electric displacement field $D$
(\prettyref{fig:Fig2}). A comparison of the three structures reveals, however, that
TQG and TPG are more tunable than TTG. This is apparent both in the $D$-dependent evolution of 
the filling range where superconductivity is measured 
(\prettyref{fig:Fig2}a-c) as well as in the critical temperature $T_c$ (\prettyref{fig:Fig2}d-f). 
Notably, superconductivity in TQG and TPG is fully 
quenched for all fillings at $D/\epsilon_0 = 0.75~\text{V nm}^{-1}$ and 
$D/\epsilon_0 = 0.6~\text{V nm}^{-1}$, respectively. In the case of TTG, 
however, superconductivity is present up to the maximum accessible electric 
field $D/\epsilon_0 = 1~\text{V nm}^{-1}$. 
Nevertheless, $R_{xx}$ versus $D$ and temperature measurements do show 
that superconductivity is suppressed at optimal doping in all three 
heterostructures; further, they reveal that $T_c$ forms a $D$ symmetric 
dome maximized at small finite 
$D$ fields (\prettyref{fig:Fig2}d-f, for electron-side data showing similar 
behaviour see \prettyref{exfig:Tc_vs_D_electron}). 
We also note that TTG, TQG, 
and TPG all exhibit a similar variation of $T_c$ when viewed as a function 
of the potential difference $U$ between the top and bottom layers 
(\prettyref{exfig:Tc_vs_D_electron}d,e; see also SI, section \ref{methods:ttg_tqg_gap} for the
energy conversion from $D$ to $U$). This layer-number invariance is consistent with
non-interacting continuum-model calculations tracking the evolution of the
inverse of the flat-band bandwidth with $U$ (\prettyref{fig:Fig2}g bottom). The
dependence of $T_c$ on $D$ in all devices qualitatively matches the predictions
of Ref.~\citenum{qinInPlaneCriticalMagnetic2021} for TTG with one marked
exception: the observed vanishing of superconductivity and the decay of $T_c$
appears to be linear in $D$ (\prettyref{fig:Fig2}e,f and
\prettyref{exfig:Tc_vs_D_electron}), in line with predictions for multilayer
graphene with rhombohedral
stacking\cite{kopninHightemperatureSurfaceSuperconductivity2011} and in contrast to
the exponential `tail’ typically expected from the weak-coupling theory (and seen in the model 
of Ref.~\citenum{qinInPlaneCriticalMagnetic2021}).

\begin{figure}[p]
    \centering
    \includegraphics[width=15cm]{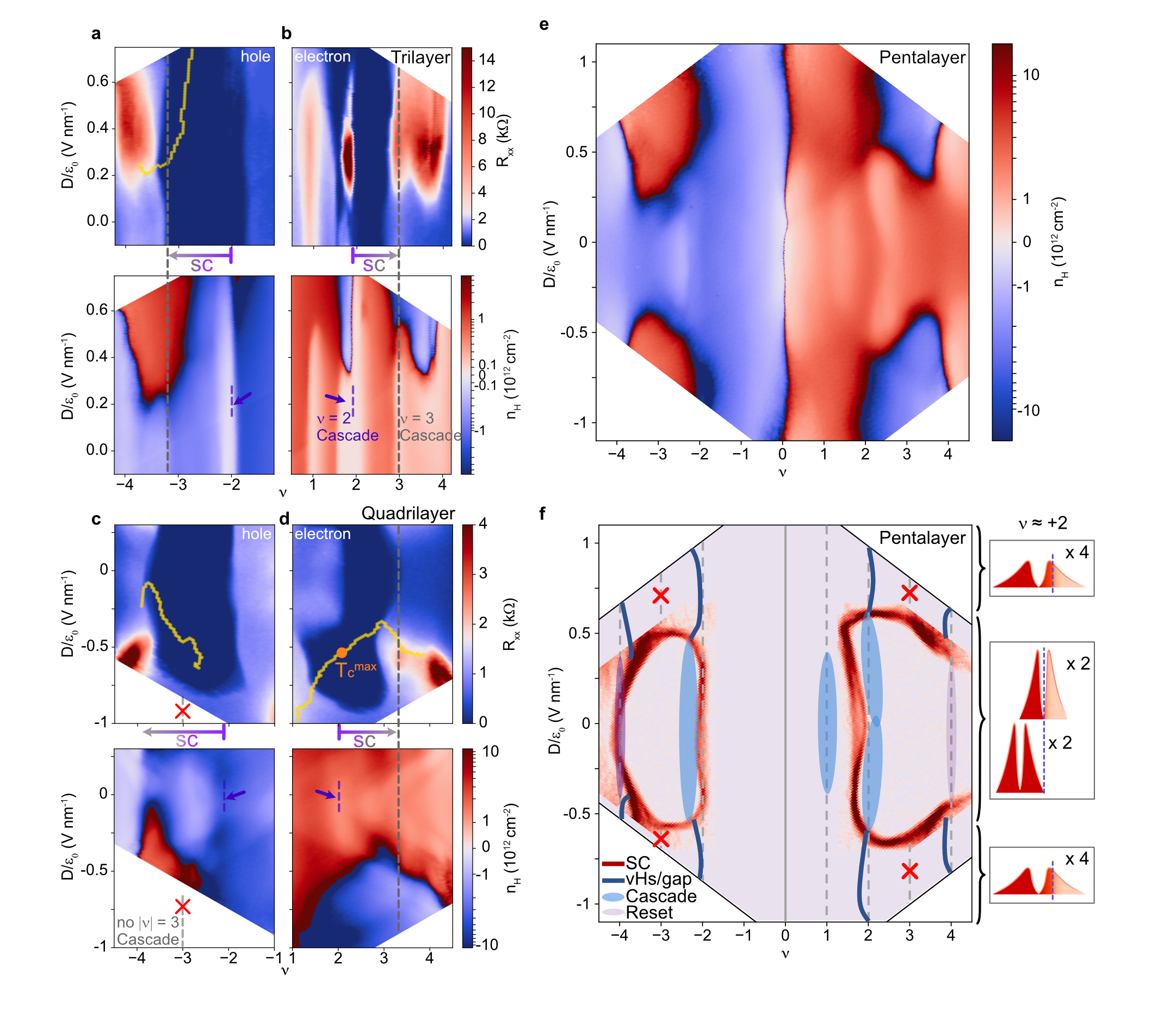}
    \caption{{\bf Interplay between superconductivity, flavour symmetry-breaking
        transitions and van Hove singularities in TTG, TQG and TPG.} 
      {\bf a},{\bf b}, $D$ field and $\nu$ dependence of $R_{xx}$ (top) and Hall
      density (bottom, measured at $B=0.9$~T) for TTG. Purple and grey dashed
      lines mark the filling factors where flavour symmetry-breaking transitions
      associated with $|\nu| = 2$ and $|\nu| = 3$ happen, respectively. The
      yellow line in {\bf a} delineates the evolution of the vHs. {\bf c},{\bf d},
      $D$ field and $\nu$ dependence of $R_{xx}$ (top) and Hall density (bottom,
      measured at $B=1.5$~T) for TQG. Superconducting $T_c$ reaches its maximum
      (orange dot in {\bf d}) exactly at the position of the vHs. 
      When present, flavour symmetry-breaking transitions around $|\nu|\approx3$ 
      coincide with the termination of superconductivity ({\bf a}, {\bf b}, {\bf d}).
      By contrast, superconductivity extends much further in the absence of
      a $|\nu| \approx 3$ reset  ({\bf c}).
      {\bf e}, $D$ field and $\nu$ dependence of
      Hall density for TPG measured at $B=1.5$~T. {\bf f}, Schematic of 
      Hall density ({\bf e}) and
      $R_{xx}$ (Fig. 2c) features for the pentalayer, including the superconducting
      boundary (red), vHs/`gap’ (dark blue), cascade (light blue), 
      and $|\nu_\mathrm{flat}|=4$ Hall density reset (light purple). 
      Sketches of the DOS around $\nu = +2$ for different $D$ fields are 
      shown on the right. The middle figure
      illustrates the flavour symmetry polarization observed in regions
      that support superconductivity. Flavour symmetry is preserved
      at higher $D$ fields, as shown in the top and bottom images. 
    }
    \label{fig:Fig3}
\end{figure}

Comparing the location of the superconducting regions with the evolution of the
Hall density as a function of $D$ and $\nu$ in TTG, TQG, and TPG provides
further insight into the intricate relationship between the superconducting
phase and the correlation-modified Fermi surface (\prettyref{fig:Fig3}). As in
previous TBG and TTG measurements, we observe symmetry-breaking electronic
transitions (a `cascade' of transitions) that are signalled by sudden drops in
the Hall density magnitude (a `reset') without a change in sign. These resets
(see dashed lines in \prettyref{fig:Fig3}a-d) indicate a rearrangement of
spin/valley sub-bands and typically occur near integer fillings of the flat
bands\cite{zondinerCascadePhaseTransitions2020,
  wongCascadeElectronicTransitions2020}. At low $D$ fields, superconducting
pockets onset around the $|\nu| = 2$ resets (purple dashed line), and the
filling extent of superconductivity varies depending on the presence or absence
of a $|\nu| = 3$ flavour symmetry-breaking transition (grey dashed line). 
For electron- and hole-doped TTG as well as for electron-doped TQG
(\prettyref{fig:Fig3}a,b,d), a flavour symmetry-breaking transition appears at
$|\nu|=3$ and superconductivity accordingly terminates. By contrast, when
signatures of the $|\nu| = 3$ reset are completely absent (for example in
hole-doped TQG, \prettyref{fig:Fig3}c, or in TPG), superconductivity extends
much further. Combined, these observations suggest that superconductivity is
favoured when only two out of the four flavours are significantly populated
($|\nu| = 2$ cascade) and suppressed beyond $|\nu|=3$ resets. This behaviour can
be understood within the simplest iteration of the cascade scenario: resets at
$|\nu|=3$ produce spin- and valley-polarized
bands\cite{potaszExactDiagonalizationMagicAngle2021,
  shavitTheoryCorrelatedInsulators2021, xieTwistedBilayerGraphene2021} 
  and naturally disfavour Cooper pairing 
  of time-reversed partners. 

At high $D$ fields, signatures of the cascade vanish and instead van Hove
singularities (vHs) become more prominent, reflecting qualitative changes in the
band structure (see yellow lines in \prettyref{fig:Fig3}a-d and
\prettyref{exfig:Hall density and vHs} that track the vHs). Consistent with
previous TTG
measurements\cite{parkTunableStronglyCoupled2021,haoElectricFieldTunable2021},
the vHs in our TTG sample (as well as in TPG, see \prettyref{fig:Fig3}e,f)
crudely bound the superconducting regions. By contrast,  the vHs in TQG
cross well into the superconducting pockets---in fact, for electron doping,
$T_c$ reaches its maximum exactly at the position of the vHs
(\prettyref{fig:Fig3}d, orange dot and \prettyref{exfig:gatemapT}d-f). The
interplay between the vHs and superconductivity is thus not a universal property
of graphene moir\'e systems but rather depends on the layer number and possibly
the precise twist angle.

Pentalayer measurements provide additional signatures that point towards a close
relation between superconducting phase boundaries and flavour
symmetry-breaking cascades (\prettyref{fig:Fig3}e,f). In contrast to TTG, in TPG
we can access $D$ fields that are large enough to stifle superconductivity---which 
occurs simultaneously with the onset of the vHs and the apparent
suppression of the cascade transitions (see red and light blue lines in
\prettyref{fig:Fig3}f that mark the superconducting boundaries and the cascade transitions,
respectively). 
For example, at low $D$ fields ($|D|/\epsilon_0 < 0.6~\text{V nm}^{-1}$) around
$\nu = +2$, the Hall density resets close to zero, in line with a nearly
complete flavour symmetry-breaking polarization. However, at higher $D$ fields
($|D|/\epsilon_0> 0.6~\text{V nm}^{-1}$), the Hall density is dominated by a vHs
around $\nu=+2$, while the cascade signatures are diminished. 
Superconductivity accordingly also vanishes. For hole doping, the disappearance of
superconductivity similarly coincides with the weakening of the cascade. This
on/off correspondence between the two phenomena suggests that they either share
a common origin, such as a large DOS, or that the cascade serves as a
prerequisite for robust superconductivity in graphene moir\'e superlattices.

\begin{figure}[p]
    \centering
    \includegraphics[width=16cm]{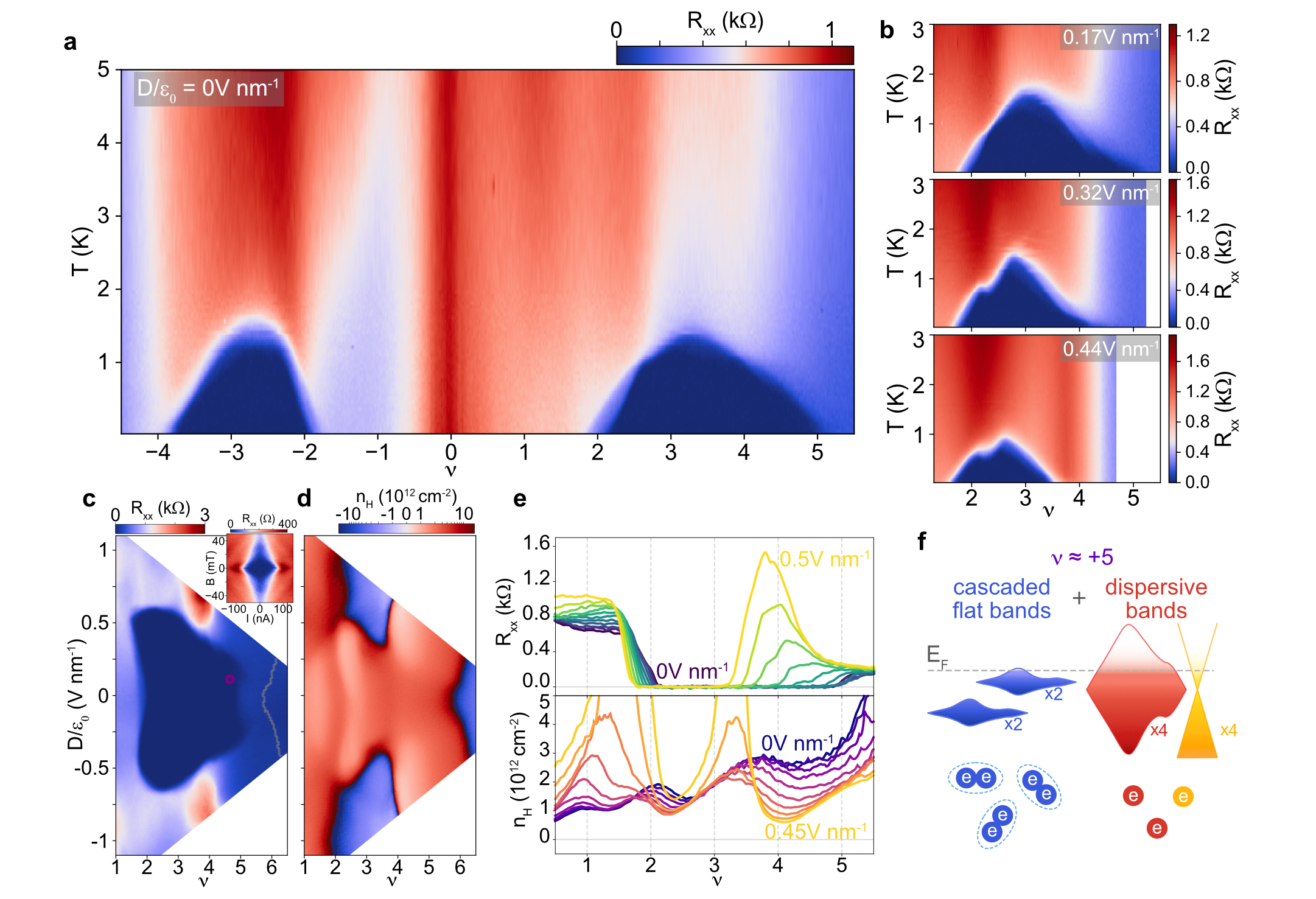}
    \caption{{\bf Extended superconducting pockets in TPG.} {\bf a}, 
    $R_{xx}$ versus $\nu$ and temperature at zero $D$
      field for twisted pentalayer graphene. {\bf b}, $R_{xx}$ versus
      temperature and $\nu$ on the electron side at $D/\epsilon_0 = 0.17$, $0.32$,
      and $0.44~\text{V nm}^{-1}$. The evolution of the
      superconducting domes and resistance peaks near $\nu=+2$ and $\nu=+4$ 
      with $D$ is shown. {\bf c},{\bf d}, $D$ field and $\nu$ dependence of
      $R_{xx}$ ({\bf c}) and Hall density ({\bf d}, measured at $B=1.5$~T),
      showing the region around the electron-side superconducting pocket. The grey line in {\bf
        c} marks the vHs originating from the dispersive TBG-like bands (see also
      \prettyref{exfig:vHS_dispersiveband}). The inset plots the Fraunhofer-like interference
      at $\nu = +4.6$, $D/\epsilon_0 = 0.12~\text{V nm}^{-1}$ (marked
      by a red dot in the main panel), confirming the robustness of the
      superconductivity above $\nu=+4$. {\bf e}, Line cuts of $R_{xx}$ (top) and Hall density (bottom, measured at $T=1.5$~K,
      $B=0.5$~T) versus $\nu$ for a range of $D$ fields (traces are shown for every
      $0.05~\text{V nm}^{-1}$ for both $R_{xx}$ and Hall density). 
      Both the presence of Hall density resets around $\nu = +4$ and the development of superconductivity extending from $\nu = +2$ to $+5$ are shown to persist for a wide range
      of $D$ fields. 
      {\bf f}, Schematic of scenario
      $(ii)$ with a Hartree correction for superconductivity at $\nu \approx +5$.
      The Hartree correction shifts the dispersive TBG- and MLG-like bands down in energy,
      which causes the flat TBG-like bands to fill 
      more slowly with doping, thus allowing them to host superconductivity at
      $\nu > +4$. }
    \label{fig:Fig4}
\end{figure}

As mentioned above, for low $D$ fields in TPG, the superconducting pockets are 
extraordinarily large, spanning $-4 \lesssim \nu < -2$ for hole doping 
and $+2 \lesssim \nu \lesssim +5$ for electron doping 
(\prettyref{fig:Fig1}e, \prettyref{fig:Fig2}c, and \prettyref{fig:Fig4}). 
In particular, the electron-side range corresponds roughly to a density 
window of $6\times 10^{12}~\text{cm}^{-2}$, which is the largest filling 
range so far reported in a graphene-based superconductor. The observed 
superconductivity exhibits similar values of $T_c$ and $B_c$ as 
the trilayer and quadrilayer samples and is likewise accompanied by a 
Fraunhofer pattern (\prettyref{fig:Fig4}c inset), confirming its robust 
nature. We emphasize that the unprecedented persistence of superconductivity across
a large filling factor range in TPG (and also TQG in comparison to TTG or
TBG) cannot be explained in a minimal framework of alternating 
twisted graphene multilayers\cite{khalafMagicAngleHierarchy2019, ledwithTBNotTB2021} 
without invoking the non-trivial role of the additional bands.

Explanations for the enlarged superconducting intervals can generically be
organized into three scenarios depending on the filling of the flat TBG-like
bands $\nu_\mathrm{flat}$, relative to the total filling $\nu_\mathrm{max}$ at
which superconductivity terminates ($\nu_\mathrm{max}=+5$ for electron-doped TPG
and $|\nu_\mathrm{max}|=4$ for TQG and hole-doped TPG). In scenario $(i)$,
$\nu_\mathrm{max}$ corresponds to $\nu_\mathrm{flat} \approx +3$, the flat-band
filling at which superconductivity is typically suppressed in TBG, whereas
in scenario $(ii)$, $\nu_\mathrm{max}$ coincides with
$\nu_\mathrm{flat}\approx +4$, precluding any simple analogy with TBG. 
Finally, scenario $(iii)$ assumes full filling of the flat bands 
\emph{before} superconductivity is suppressed at $\nu_\mathrm{max}$.
This scenario includes the possibility that the distinction between the 
different TBG- and MLG-like bands breaks down even at $D=0$ due to 
hybridization (for a more complete description of the three scenarios, see SI, 
section~\ref{methods:scenario}).

From the perspective of the non-interacting band structure, scenarios $(i)$ and
$(ii)$ are completely implausible. In particular, although the presence of the
dispersive bands implies that $\delta\nu=|\nu|-|\nu_\mathrm{flat}|>0$, this
effect is much smaller than needed for these two scenarios. However, $\delta\nu$
may nevertheless be significantly enhanced by Coulomb interactions. First,  
the Hartree correction accounts for the system's desire for a spatially uniform 
charge distribution. Since the flat bands are highly localized on the AA sites, 
the Hartree correction, so far primarily studied in TBG, manifests mainly as a 
band deformation and
flattening\cite{guineaElectrostaticEffectsBand2018,
  rademakerChargeSmootheningBand2019, goodwinHartreeTheoryCalculations2020,
  calderonInteractions8orbitalModel2020,
  choiInteractiondrivenBandFlattening2021}---shifting the density of states to
spread out the charge. When dispersive bands, whose wavefunctions are more
uniformly distributed within the unit cell, are also present as in TTG, TQG, and
TPG, charge may be redistributed by shifting the energy of these bands
relative to the flat bands\cite{fischerUnconventionalSuperconductivityMagicAngle2021,
  kimSpectroscopicSignaturesStrong2021}. More generally, the Coulomb interaction 
  can facilitate symmetry breaking, as reflected in the flat-band cascade 
  resets and gap openings (which in TBG yields correlated insulators\cite{caoCorrelatedInsulatorBehaviour2018}). In this context, gap formation pushes the flat bands up in energy, allowing 
  additional charge to accumulate in the dispersive bands, thus further increasing $\delta\nu$.
Finally, multilayer structures beyond TBG can additionally have non-uniform layer-to-layer 
charge distribution or next-layer coupling which may further deform the bands, leading to 
self-generated shifts between the flat and the other bands as well as introducing coupling
between them (see SI, sections \ref{methods:band_mixing} and \ref{methods:interlayer_chem_potential}).

A toy model for TPG incorporating these mechanisms (see SI,
section~\ref{method:interactions_in_tpg}) suggests a minimal flat-band occupation
$\nu_\mathrm{flat}\gtrsim+3.8$ at $\nu\approx+5$, diminishing the plausibility
of scenario $(i)$ for electron-doped TPG which has $\nu_\mathrm{max}\approx+5$.
The relevance of this scenario is further undermined by the observation of
vHs at $\nu\approx+6$ (\prettyref{exfig:vHS_dispersiveband}d):
under the reasonable assumption that the non-interacting band structure remains
valid for the dispersive TBG-like bands (apart from a Hartree shift), scenario $(i)$ would
instead place the observed vHs near $\nu\approx+5$. Taken together, these arguments 
effectively rule out scenario $(i)$. Note, however, that the presented
line of reasoning is not straightforward for the other superconducting pockets 
(see SI, section \ref{methods:scenario}).

Both scenarios $(ii)$ and $(iii)$ are indicative of the non-trivial role of
additional bands in stabilizing superconductivity. Assuming well-defined flat 
and dispersive bands, in scenario $(iii)$ the former bands are completely filled, 
and superconductivity is supported fully by the latter non-flat bands. This assertion 
is at odds with the large dispersion of the remaining TBG- and MLG-like bands.  
However, while the exact mechanism underlying scenario $(iii)$ is difficult 
to pin down, it is not without experimental support. For instance, a natural 
interpretation of the Hall density minimum around $\nu\approx+4$ for 
$|D|\lesssim0.4\,\text{V nm}^{-1}$ is that it marks the complete filling 
of the flat bands, $\nu_\mathrm{flat}\approx+4$ 
(\prettyref{fig:Fig4}e and \prettyref{exfig:Hall_density_TPG};
see also SI, section \ref{methods:scenario} for more discussion). 

One possible realization of scenario $(iii)$ consistent with the experimental 
observations is that the division of the electronic states into simple 
TBG- and MLG-like bands fails completely---obviating our very definition of 
$\nu_\mathrm{flat}$ and potentially allowing flavour 
polarization, and accompanying superconductivity, to persist well beyond 
$\nu=+4$. While such hybridization is expected for finite $D$ fields, 
mixing between flat, dispersive TBG- and MLG-like bands for $|\nu|<|\nu_\mathrm{max}|$ 
may occur even at $D=0$. For example, hybridization could result from 
mirror symmetry breaking due to interactions or proximity to WSe$_2$. 
Importantly, in TQG and TPG even terms that preserve mirror symmetry, such as 
layer-to-layer charge inhomogeneity or distant-layer coupling, allow for 
band hybridization (see SI, sections \ref{methods:band_mixing} and 
\ref{methods:interlayer_chem_potential}). 
This feature distinguishes 
TPG and TQG structures from TTG and may therefore play a role in explaining  
extensive superconducting regions. Finally, we mention that other effects,
such as strain\cite{biDesigningFlatBands2019}, or a different stacking order, 
may yield multiple sets of flat bands\cite{xieAlternatingTwistedMutilayer2021} 
(see SI, section \ref{methods:stacking}) even at the non-interacting level, in which case 
multiple bands can host superconductivity independently. 
Importantly, however, invoking this explanation would place TQG and TPG well outside a 
simple TBG paradigm, as coexisting but independent sets of flat TBG-like bands 
are expected to produce more cascade resets than observed experimentally, and therefore 
are unlikely. 

Our measurements demonstrate the increasing predominance of superconductivity in
twisted gra\-phene multilayer structures as the number of layers is increased from
three to five and highlight the close relationship between the flavour
symmetry-breaking transitions and superconductivity.  Moreover, our findings 
suggest a scenario in which the symmetry-broken $\nu=\pm2$
state strongly favours the formation of the superconducting state while the
cascade corresponding to $\nu=\pm3$ suppresses it. Interestingly, this scenario 
is consistent not only with previous TBG observations but also in part with the 
recently investigated ABC trilayers\cite{zhouSuperconductivityRhombohedralTrilayer2021} 
and Bernal bilayers\cite{zhouIsospinMagnetismSpintriplet2021} where superconductivity is 
observed near symmetry-breaking transitions. This universality 
appears to suggest a possibility that superconductivity in graphene-based superconductors 
originates from a common underlying symmetry-broken state. In this context, our 
discovery of superconductivity in TQG and TPG together with recent work on 
untwisted bi- and trilayers dramatically expands the scope of graphene-based 
superconductors. This expansion holds promise for resolving important questions 
related to the nature of the pairing mechanism in these systems and provides 
guidance for developing novel graphene-based superconductors and their 
applications. 

\section*{Methods}

\textbf{Device fabrication:} All devices were fabricated using a `cut and stack’
method, in which graphe-ne flakes were separated into pieces using a
sharp tip (made out Platinum-Iridium); this approach prevents unwanted twisting 
and strain during tearing while allowing more control over the flake 
size and shape. After cutting, stacking procedure was as follows: first, a thin 
hBN flake ($10-30$~nm) is picked up using a propylene carbonate (PC) film
previously placed on a polydimethylsiloxane (PDMS) stamp. Then the hBN flake is used
to pick up an exfoliated monolayer of WSe$_2$ (commercial source, HQ graphene)
before approaching the graphene. After picking up the first piece of the graphene
flake, the following layers are twisted by an angle $\pm\theta$ relative to the
previous one in an alternating sequence. Transfer stage rotation $\theta$
overshoots the target angle by $0.1-0.2\degree$ to construct the measured angles. 
Care was taken to approach and pick up each stacking step slowly.
In the last step, a thicker hBN ($30-70$~nm) is picked up, and the whole stack
is dropped on a predefined local gold back gate at $150\degree$C while the PC is
released at $170\degree$C. The PC is then cleaned off with
N-Methyl-2-Pyrrolidinone (NMP). The final geometry is defined by dry etching
with a CHF$_3$/O$_2$ plasma and deposition of ohmic edge contacts (Ti/Au, 5
nm/100 nm) and top gate.

\textbf{Measurements:} All measurements were performed in a dilution
refrigerator (Oxford Triton) with a base temperature of $\sim25$~mK, using
standard low-frequency lock-in amplifier techniques. Unless otherwise specified,
measurements are taken at the base temperature. Frequencies of the lock-in
amplifiers (Stanford Research, models 830 and 865a) were kept in the range of
$7-20$~Hz in order to measure the device's DC properties and the AC excitation
was kept $<5$~nA (most measurements were taken at $0.5-1$~nA to preserve the
linearity of the system and avoid disturbing the fragile states at low
temperatures). Each of the DC fridge lines pass through cold filters, including
4 Pi filters that filter out a range from $\sim80$~MHz to $>10$~GHz, as well as
a two-pole RC low-pass filter.

\noindent {\bf Acknowledgments:} We thank
Haoxin Zhou and Soudabeh Mashahadi for fruitful discussions. 
{\bf Funding:} This work has 
been primarily supported by NSF-CAREER award (DMR-1753306), and 
Office of Naval Research (grant no. N142112635), and Army Research 
Office under Grant Award W911NF17-1-0323. Nanofabrication efforts 
have been in part supported by Department of Energy DOE-QIS program 
(DE-SC0019166). S.N-P. acknowledges support from 
the Sloan Foundation (grant no. FG-2020-13716). G.R., J.A., 
and S.N.-P. also acknowledge support of the Institute for 
Quantum Information and Matter, an NSF Physics Frontiers 
Center with support of the Gordon and Betty Moore Foundation 
through Grant GBMF1250. C.L. acknowledges support from the 
Gordon and Betty Moore Foundation’s EPiQS Initiative, 
grant GBMF8682. Y.P. acknowledges support from the startup 
fund from California State University, Northridge. F.v.O. is supported
by CRC 183 (project C02) of Deutsche Forschungsgemeinschaft.

\noindent {\bf Author Contribution:} Y.Z. and R.P. performed the measurements,
fabricated the devices, and analyzed the data. Y.C. and H.K. helped with device 
fabrication and data analysis. C.L., A.T. and Y.P. developed 
theoretical models and performed calculations in close collaboration and guidance by 
F.v.O., G.R. and J.A. K.W. and T.T. provides hBN crystals. S.N-P. 
supervised the project. Y.Z., R.P., C.L., A.T., Y.P., F.v.O., G.R., J.A., 
and S.N-P. wrote the manuscript with the input of other authors. 

\noindent{\bf Competing interests:} The authors declare no competing interests.

\noindent {\bf Data availability:} The data supporting the findings of this
study are available from the corresponding authors on reasonable
request.

\noindent {\bf Code availability:} All code used in modeling in this
study is available from the corresponding authors on reasonable
request.

\clearpage

{\large {\bf Supplementary Information:}} \\ 
{\large Ascendance of
Superconductivity in Magic-Angle Graphene Multilayers} \\
Yiran Zhang, Robert Polski, Cyprian Lewandowski, 
Alex Thomson, Yang Peng, Youngjoon Choi, 
Hyunjin Kim, Kenji Watanabe, Takashi Taniguch, 
Jason Alicea, Felix von Oppen, Gil Refael, 
and Stevan Nadj-Perge

\section{Device Uniformity and Twist Angle Assignment}
\label{methods:uniformity}
\begin{figure}[p]
    \includegraphics[width=16cm]{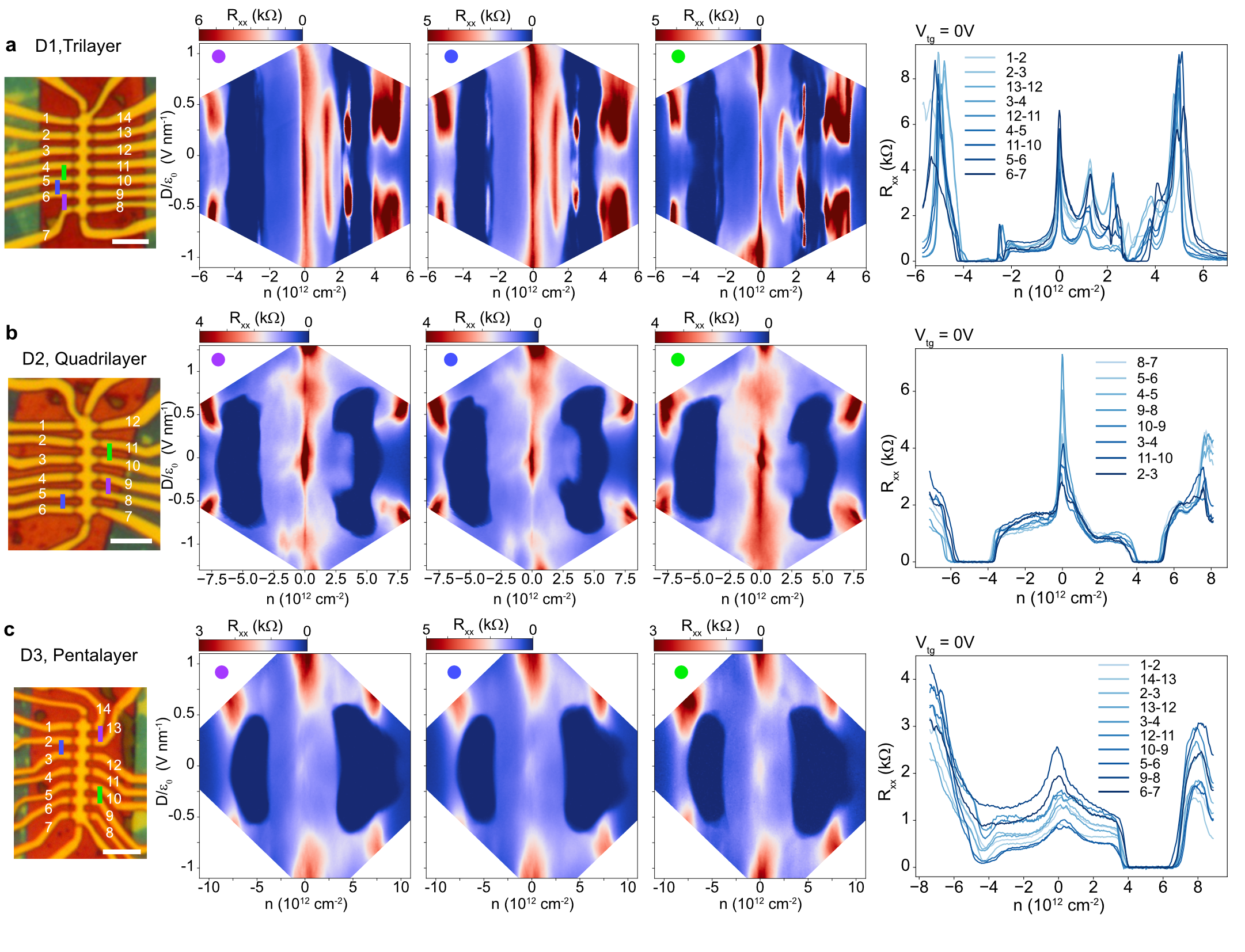}
    \centering
    \caption{{\bf Sample uniformity and reproducibility of the results.} {\bf a}--{\bf c},
      Leftmost optical images are D1--D3 mentioned in the main text. The scale
      bar in each panel corresponds to $5~\mu$m. $R_{xx}$ versus density and displacement field ($n$--$D$) plots shown 
      in the middle are obtained from electrodes marked with the corresponding colored lines. The electrodes marked with purple lines were used for measuring $R_{xx}$ in
      the main text. Rightmost plots are $R_{xx}$ versus carrier density with
      top-gate voltage fixed at V$_\textsubscript{tg}$ = 0~V. All three devices
      have a high degree of homogeneity in twist angle with the same
      superconducting filling range and $|\nu| = 4$ carrier density for multiple
      contacts. The behaviour of superconductivity and other symmetry-breaking
      features is highly reproducible for different contacts. }
\label{exfig:uniformity}
\end{figure}

\textbf{Device homogeneity and effect of WSe$_2$:} All
devices investigated here show a high degree of twist angle homogeneity as characterized 
by four-point measurements between different pairs of contacts.
\prettyref{exfig:uniformity} shows $R_{xx}$ versus carrier density with fixed top-gate
voltage (V$_\textsubscript{tg}$ = 0~V), revealing that almost every pair of contacts shows
superconductivity. More importantly, superconducting pockets from different
pairs significantly overlap in the filling range, and resistance peaks at $|\nu| = 4$
appear at the same density. Moreover, all findings related to the extent of the superconducting 
phase and the occurrence of the symmetry-breaking transitions in the $\nu$--$D$ phase diagram 
are highly reproducible. This also includes the observation of a gapped correlated 
insulator at $\nu = +2$ in TTG, which has not been reported previously. In this context,
we note that any significant twist-angle disorder would create conducting percolation 
pathways that quickly suppress insulating behaviour. 

We attribute the low level of disorder to the use of monolayer WSe$_2$ during device
stacking, presumably originating from the increased lateral friction between 
WSe$_2$ and graphene (compared to the friction at the hBN-graphene interface). 
We note that this additional layer does not change the magic-angle 
condition\cite{aroraSuperconductivityMetallicTwisted2020, 
choiCorrelationdrivenTopologicalPhases2021}, and the induced spin-orbit 
interaction (SOI) energy scale is $\sim1$~meV in twisted 
bilayers\cite{aroraSuperconductivityMetallicTwisted2020}. Therefore, SOI is 
likely too small to significantly affect the overall band structure and directly impact 
the cascade physics. Finally, we note that, in general, SOI is expected to 
manifest differently when the sign of $D$ field is reversed, a feature that has not been 
observed in the experiment. 
 
\textbf{Twist angle assignment in multilayers:} Twist angles
were determined from high $B$ field data and corresponding Landau-fan diagrams 
in a similar way as in TBG. From the slope of the Landau fan at charge neutrality 
(which is directly proportional to the gate-sample capacitance) and the voltage difference 
between charge-neutrality point (CNP) and $|\nu|=4$ filling, the corresponding $|\nu| = 4$ electron density 
is obtained. We used two separate criteria for the assignment of $|\nu|$ = 4. 
First, at high $D$ fields, resistive features (peaks) emerge (\prettyref{fig:Fig2}a-c).  
We interpret these peaks presumably as the opening of the hybridization 
gaps and corresponding full filling ($|\nu| = 4$) of the `gapped' bands. Second, 
at high $B$ fields, quantum Hall insulating states develop around $|\nu| = 4$, which
typically cover a broader filling range where Hall conductance is quantized 
in accordance with the expectations from the dispersive bands (\prettyref{exfig:Fan diagram},
also see discussion below). Electron density of $|\nu| = 4$ directly
determines the twist angle in the low-angle approximation $\theta^2 \approx \sqrt{3}a^{2}n_{|\nu| = 4}/8$, 
where $a = 0.246$~nm is the graphene lattice constant. We note that signatures of 
the dispersive bands are also observed in Landau-fan diagrams (\prettyref{exfig:Fan diagram}). 
For example, emerging Landau levels from the dispersive bands are typically observed through 
oscillations at low magnetic field. Since at low energies, the dispersive bands (\prettyref{fig:Fig1}b
left) can be effectively treated as decoupled MLG-like bands when
considering the Landau level spectrum\cite{caoSuperlatticeInducedInsulatingStates2016}, and
the corresponding Hall conductance around $|\nu| = 4$ will be quantized in a way that depends on 
the number of layers (see \prettyref{exfig:Fan diagram}g-m for Hall conductance line cuts). 

\section{Determining $\mathbf{T_c}$ and Hall density}
\label{methods:T_B_dependence}
\textbf{$\mathbf{T_c}$ and the coherence length:} $T_c$ is determined by the following 
procedures. First, the high temperature $R_{xx}$ data is fitted using a linear function 
$R(T) = AT + B$. Then, $T_c$ is defined by the value where $R_{xx}(T)$ is a certain fraction
(typically $10\%$ as in \prettyref{fig:Fig2}) of $R(T)$. Ginzburg-Landau coherence 
lengths $\xi_\mathrm{GL}$ are obtained from the $B$ dependence of $T_c$, by fitting the Ginzburg-Landau relation 
$T_c/T_{c0}=1-(2\pi\xi_\mathrm{GL}^2/\Phi_0)B_\perp$, where $\Phi_0=h/(2e)$ is the 
superconducting flux quantum and $T_{c0}$ is the critical temperature at zero 
magnetic field. We get $\xi_\mathrm{GL}$ from the $T_c$ vs. $B$ linear fit, where the intercept at the $B$ axis is equal to $\Phi_0/(2\pi\xi_\mathrm{GL}^2)$. Following Ref.~\citenum{parkTunableStronglyCoupled2021}, we use $T_{c}$ defined by $40\%$ of the normal state resistance to evaluate the coherence 
length data in \prettyref{exfig:Fraunhofer}e (corresponding error bars are evaluated by using $T_{c}$ defined 
by $30\%$ and $50\%$ of the normal state resistance). As mentioned in the main text,
$\xi_\mathrm{GL}$ ($B_c$) is much smaller (higher) in the twisted graphene
multilayers compared to TBG. One possibility for the reduction of $\xi_\mathrm{GL}$
is the relative decrease of the characteristic moir\'e wavelength 
(see \prettyref{exfig:Fraunhofer}f).

\textbf{Hall density analysis:} Hall density shown in \prettyref{fig:Fig3} is
obtained by converting the anti-symmetric part of the $R_{xy}$ data, i.e., by subtracting 
data measured at positive and negative magnetic fields. We used either $|B| = 0.9$~T or 
$1.5$~T in order to fully suppress superconductivity. Previously, it was found that 
in TTG\cite{parkTunableStronglyCoupled2021}, at high $D$ fields superconductivity is 
bounded by regions corresponding to vHs, i.e., when Hall density changes sign. We approximately 
find a similar trend in our TTG and TPG structures, although vHs occasionally intrude superconducting pockets slightly. We note that the exact positions of vHs 
depend on the precise magnetic field used in the measurements 
(for example, see \prettyref{exfig:Hall_density_TPG}a and c); however, this 
effect is relatively small relative to the observed intrusions. More importantly, 
TQG behaviour is qualitatively different, as we find that positions of vHs 
and boundaries of superconducting pockets are independent. We also note that 
resets associated with the flavour polarization do not move in the $B$ fields 
($B\approx1$T) used to extract Hall density evolution. The occasional shift
of these resets from integer $\nu$ values, may be attributed to either
effects of interactions (i.e. Hartree correction, see section \ref{methods:Hartree}) 
or the details of cascade physics\cite{zondinerCascadePhaseTransitions2020} at finite temperatures\cite{saitoIsospinPomeranchukEffect2021,rozenEntropicEvidencePomeranchuk2021}.

\section{Insulating Behaviour in TTG and TQG}
\label{methods:ttg_tqg_gap}
\textbf{$\mathbf{\nu = +2}$ correlated insulators in TTG:} 
In our TTG device, we observed a previously unidentified $\nu = +2$ correlated 
insulator state (\prettyref{fig:Fig1}f and \prettyref{fig:Fig1}c inset). This state is highly sensitive to 
$D$ field with maximum activation gap reaching $\Delta_{+2}=0.27$~meV 
(see \prettyref{exfig:CI_TTG} for detailed $D$ and $\nu$ dependence). Also, both in-plane 
and out-of-plane $B$ field suppress the insulating behaviour. These experimental 
observations are highly indicative of a gap that originates from strong 
interactions in TTG. Since it is pinned to $\nu = +2$, it likely shares 
the same underlying origin as the TBG gap found at 
half filling\cite{caoCorrelatedInsulatorBehaviour2018}. We note, however, that 
formation of the fully gapped states in TTG requires a mechanism that
additionally gaps out the MLG-like band, which may explain the presence of the gap only at finite
$D$ fields. Moreover, suppression of the gap with magnetic 
field is at odds with the $C_2$ breaking scenario\cite{serlinIntrinsicQuantizedAnomalous2019,
  sharpeEmergentFerromagnetismThreequarters2019} and is more in line with 
  incommensurate Kekul\'e spiral\cite{kwanKekulSpiralOrder2021} or
intervalley-coherent\cite{bultinckGroundStateHidden2020, lianTwistedBilayerGraphene2021,
xieTwistedSymmetricTrilayer2021, christosCorrelatedInsulatorsSemimetals2021} orders in the flat bands.  
Finally, we can in part rule out that the gap originates from induced SOI. For example, 
terms corresponding to Rashba SOI change sign upon $D$-field inversion, yet experimentally we
find similar gap values for both positive and negative $D$ fields. However, it is still possible 
that SOI promotes instabilities that favour the formation for certain $\nu = +2$ insulating states in TTG. 
Future work will address the nature of this state in more detail.

\textbf{Charge-neutrality gaps in TQG and conversion between $U$ and $D$:}
\prettyref{exfig:CI_TTG}f and g shows the charge-neutrality gap of TQG as a function of
$D$ field or potential difference $U$ (between the top and the bottom
graphene layer). From the continuum model, a gap in TQG is expected when finite 
$D$ field is applied. However, the details of the gap evolution depend on the precise 
twist angle. When the twist angle is below the magic-angle value, a charge-neutrality 
gap opens as soon as a finite $D$ field is applied. On the other hand, when the twist angle 
is above the magic-angle value, a gap opens only at much higher $D$ fields. 
The gap opening at $D/\epsilon_0 \approx 1.1~\text{V nm}^{-1}$ in our TQG structure 
is consistent with the device being slightly above the magic angle. Note that the 
charge-neutrality gap is a good reference for matching the experimental $D$ field with 
the potential difference $U$ used in calculations since the interaction-driven Hartree 
correction vanishes at CNP. For a direct comparison, we enforce a more realistic flat-band 
bandwidth of $\sim20$~meV in the continuum model by slightly tuning away from the magic
angle, and get a $U$-dependent gap size (\prettyref{exfig:CI_TTG}g). A good match 
between the experimental and the calculated gap is found when converting
$D$ into $U$ with an empirical factor:
$U = 0.1\times (n-1)\times0.33~{\rm nm}\times eD$, where $e$ is the electron
charge and $n-1$ is the number of graphene interfaces. This conversion
is used for the other parts of the paper, for example, $T_c$ versus $U$ in
\prettyref{exfig:Tc_vs_D_electron}. We note, however, that relative comparison 
(i.e. scaling) between TTG, TQG, and TPG (in the context of $T_c$) does not rely 
on the precise $D$ to $U$ conversion.
  
\newcounter{theorysubsection}
\renewcommand{\thetheorysubsection}{\thesection\alph{theorysubsection}}
\newcommand{\theorysubsection}[1]{\refstepcounter{theorysubsection}
\subsection{\alph{theorysubsection}. #1:}}
\newcommand{\theorysubsubsection}[1]{\textit{\textbf{#1}}}

\section{Theoretical Calculations}
\label{methods:calculations}

In this section, we describe the non-interacting continuum model for 
multilayers and how symmetry considerations and various interaction terms 
affect the band structure of TTG, TQG, and TPG.

\theorysubsection{Continuum model}
\label{methods: continuum_model}

Band structure calculations are performed using a straightforward generalization 
of the TBG continuum
model\cite{lopesdossantosGrapheneBilayerTwist2007,bistritzerMoireBandsTwisted2011}
extended to multilayer graphene
systems\cite{khalafMagicAngleHierarchy2019, carrUltraheavyUltrarelativisticDirac2020,
calugaruTwistedSymmetricTrilayer2021, leiMirrorSymmetryBreaking2021}. 
As discussed above, we consider graphene multilayer systems with 
$n_\mathrm{layer}=3,4,$ and $5$ layers ($n$ in the main text) in which the graphene sheets are 
twisted by alternating angles. In particular, we can envision grouping the 
layers into even and odd sets and then rigidly twisting these two groups 
by the twist angle $\theta$; equivalently, each layer $\ell=1,\dots,n_\mathrm{layer}$ 
is twisted by an angle $\theta_\ell=(-1)^\ell \theta/2$.
For the moment, we focus on the case where the layers are all AA stack (\emph{i.e.} 
stacked directly on top of one another) prior to twisting (see below, 
section~\ref{methods:stacking}).

It is appropriate to approximate the dispersion of the underlying graphene 
monolayers with the two Dirac cones about the valleys at $K$ and $K'$. 
Note that because of the twist, the Dirac cones are located at \emph{slightly} 
different momenta depending on whether the layer $\ell$ is even or odd, and we 
have denoted the Dirac cones' momenta here as $\vK_\ell$ and $\vK_\ell'$.
We thus define the spinors $\psi_{\ell,K^{(\prime)}}$ in terms of the 
microscopic graphene operators via 
$f_\ell(\vr)=e^{i\vK_\ell\cdot\vr}\psi_{\ell,K}(\vr)+e^{i\vK_\ell'\cdot\vr}\psi_{\ell,K'}(\vr)$.
Equivalently, in momentum space we can write 
$\psi_{\ell,K^{(\prime)}}(\vk)=f_\ell(\vk+\vK_\ell^{(\prime)})$ provided $\vk$ 
is sufficiently close to $\vK_\ell^{(\prime)}$.
In our definition of $\psi_\ell$ (and $f_\ell$) both an A/B sublattice 
index and a spin index have been suppressed.
Importantly, the small twist angle only mediates a very small momentum exchange 
between the neighbouring layers so that states originating proximate to one 
valley do not mix with those originating proximate to the other. 
We thus focus for the moment on valley $K$ and suppress the valley subscript 
until mentioned otherwise, $\psi_{\ell,K}\to \psi_\ell$.

The band structure model can be separated into a sum of two parts:
$H_\mathrm{cont}=H_\mathrm{D}+H_\mathrm{tun}$.
The first term, $H_\mathrm{D}$ is the intralayer Dirac term:
\begin{align}\label{eqn:dirac_def}
    H_\mathrm{D}
    &=
    \sum_{\ell=1}^{n_\mathrm{layer}}\int d^2\vr \,\psi^\dag_\ell(\vr) h_{\mathrm{D},\ell}(\vr) \psi_\ell(\vr),
    &
    h_{\mathrm{D},\ell}(\vr)
    &=
    v_0 e^{i\theta_\ell\sigma^z}i\big( \partial_x \sigma^x + \partial_y \sigma^y \big),
\end{align}
Here, $v_0\sim 10^6\,\text{m/s}$ is the Fermi velocity of the Dirac cones 
and $\sigma^{x,y,z}$ are Pauli matrices acting on the A/B sublattice indices 
of the spinors $\psi_\ell$. In our simulations, we assume that the 
Fermi velocity of the graphene monolayers does not differ layer to layer. 
Note that this assumption, specifically does not take into account effects 
such as graphene velocity renormalization that can occur in the top layer
due to tunnelling between the graphene monolayer and the WSe$_2$ substrate, as we do 
not expect these effects to be large enough to have appreciable impact on the 
resulting band structure.

We assume that tunnelling only occurs between adjacent layers and that it takes the form
\begin{align}
    H_\mathrm{tun}
    &=
   \sum_{\ell=1}^{n_\mathrm{layer}-1}\int d^2\vr\,\psi^\dag_\ell(\vr) T_{\ell,\ell+1}(\vr)\psi_{\ell+1}(\vr) +h.c.
\end{align}
where
\begin{align}
    T_{\ell,\ell+1}(\vr)
    &=
    \sum_{j=1,2,3}
    e^{-(-1)^\ell i\vq_j\cdot\vr} t_j,
    \notag\\
    \vq_j&=
    \frac{4\pi}{3L_M}R\left(\frac{2\pi}{3}(j-1)\right) \begin{pmatrix}0\\-1\end{pmatrix},
    \notag\\
    t_j&= w'+w\left( e^{-2\pi(j-1)i/3}\sigma^+ + e^{-2\pi(j-1)i/3}\sigma^-\right).
\end{align}
Here, $R(\phi)=e^{-i\phi \sigma^y}$ is a $2\times2$ rotation matrix acting on vector indices, $L_M=a/[2\sin(\theta/2)]$ is the moiré lattice constant, and $\sigma^\pm=(\sigma^x\pm i\sigma^y)/2$ act on the sublattice indices.
The parameters $w'$ and $w$ set the interlayer tunnelling strength; we discuss their values below.
It will be convenient to define the dimensionless ratios
\begin{align}
    \label{eqn:cont_mod_ratios}
    \eta&=\frac{w'}{w}\cCom
    &
    \alpha&=\frac{w}{v_0 k_\theta}\cCom
\end{align}
where $k_\theta=4\pi/(3L_M)=2\sin(\theta/2)\cdot4\pi/(3a)$.
The total Hamiltonian may be written in matrix form as
\begin{align}
\label{eq:total_hamiltonian}
    H_{Tn_\mathrm{layer}G}
    &=H_\mathrm{D}+H_\mathrm{tun}=
    \sum_{\ell,\ell'=1}^{n_\mathrm{layer}}\int d^2\vr\,\psi^\dag_\ell(\vr)\left[h_\mathrm{cont}(\vr)\right]_{\ell,\ell'}\psi_{\ell'}(\vr)
    \notag\\
    h_{Tn_\mathrm{layer}G}(\vr)
    &=
    \begin{pmatrix}
    h_{\mathrm{D},1}(\vr) &   T_{1,2}(\vr)  &  0    & \hdots \\
    T^\dag_{1,2}(\vr) & h_{\mathrm{D},2}(\vr)  &   T_{2,3}(\vr) &  \hdots \\
    0   &   T_{2,3}^\dag(\vr)  &   h_{\mathrm{D},3}(\vr)    &     \hdots  \\
    \vdots  &   \vdots  &   \vdots  &   \ddots
    \end{pmatrix}
\end{align}
As currently written, the diagonal Dirac terms, $h_{\mathrm{D},\ell}(\vr)$, as well as the 
off-diagonal tunnelling terms, $T_{\ell,\ell'}(\vr)$, depend only on whether $\ell$ is even or odd.
We can thus simplify the above expression by writing the Dirac terms as $h_{\mathrm{D},2\ell-1}(\vr)=h_{\mathrm{D},1}(\vr)$,
$h_{\mathrm{D},2\ell}(\vr)=h_{\mathrm{D},2}(\vr)$ and the tunnelling terms as $T_{2\ell-1,2\ell}(\vr)=T(\vr)$, $T_{2\ell,2\ell+1}(\vr)=T^\dag(\vr)$.

It has been shown\cite{khalafMagicAngleHierarchy2019} that a block diagonal form exists 
for Hamiltonians of the form Eq.~\eqref{eq:total_hamiltonian}. We provide the specific 
transformations used for three, four, and five layers below.

{\theorysubsubsection{{Twisted bilayer graphene}}}
Since the spectrum of the twisted multilayers breaks into independent 
sets of TBG- and MLG-like bands, we first briefly review the Hamiltonian 
of TBG. Thus, we start with
\begin{align}
    h_{\alpha,\eta,\theta}(\vr)
    &\equiv
    h_\mathrm{TBG}(\vr)
    =
    \begin{pmatrix}
        h_{\mathrm{D},1}(\vr)   &   T(\vr)  \\
        T^\dag(\vr) &   h_{\mathrm{D},2}(\vr)
    \end{pmatrix}.
\end{align}
Provided that inversion and time reversal symmetries are preserved, the Dirac 
cones described by $h_{\mathrm{D},\ell}(\vr)$ at $\vK_\ell$ are preserved 
even when the interlayer tunnelling is added. Nevertheless, this tunnelling 
term breaks the (effective) continuous translation symmetry of $h_{\mathrm{D},\ell}$.
Consequently, the set of conserved momenta are confined to reduced moir\' e Brillouin zone (BZ).
Like the original monolayer graphene BZ, the moir\' e BZ forms a hexagon with 
the Dirac cones located at the corners. Here, we define $\vK_1=\vec{\kappa}$ 
and $\vK_2=\vec{\kappa}'$ (for the other valley, $\vK_1'=\vec{\kappa}'$, $\vK_2'=\vec{\kappa}$).

For small twist angles, the intralayer Dirac terms are nearly identical, 
$h_{\mathrm{D},\ell}(\vr)\approx h_\mathrm{D}(\vr)=v_0(i\partial_x\sigma^x+i\partial_y\sigma^y)$---namely, 
the rotation in Eq.~\eqref{eqn:dirac_def} may be neglected to first order.
In this case, the spectrum depends solely on the ratios $\eta=w'/w$ and 
$\alpha=w/(\hbar v_0 k_\theta)$, where $k_\theta=4\pi/(3L_M)$ is the distance 
separating $\kappa$ and $\kappa'$. Further, as shown in Ref.~\citenum{bistritzerMoireBandsTwisted2011}, 
the spectrum close to Dirac points at $\kappa$ and $\kappa'$ can be approximated 
using a simple perturbative scheme. In particular, in momentum space one finds 
\begin{align}\label{eqn:vflat_approx}
    h_\mathrm{fl}(\vk+\vec{\kappa}^{(\prime)})
    &\approx
    v_{\alpha,\eta} (k_x\sigma^x + k^y\sigma^y),
    &
    v_{\alpha,\eta} 
    &=
    \frac{1-3\alpha^2}{1+3\alpha^2(1+\eta^2)}v_0.
\end{align}
The magic angle is defined\cite{bistritzerMoireBandsTwisted2011} by the 
condition $v_{\alpha,\eta}=0$, which we see here should occur for $\alpha\approx 1/\sqrt{3}$.

{\theorysubsubsection{{Twisted trilayer graphene}}}

The Hamiltonian for the three layer system is
\begin{align}
    h_\mathrm{TTG}(\vr)
    &=
    \begin{pmatrix}
    h_{\mathrm{D},1}(\vr) &   T(\vr)  &  0   \\
    T^\dag(\vr) & h_{\mathrm{D},2}(\vr)  &   T^\dag(\vr) \\
    0   &   T(\vr)  & h_{\mathrm{D},1}(\vr)
    \end{pmatrix}.
\end{align}
It maybe be transformed into a block diagonal form as
\begin{align}\label{eqn:ttg_blk_diag}
    \tilde{h}_\mathrm{TTG}(\vr)
    &=
    V_\mathrm{TTG}^\dag h_\mathrm{TTG}(\vr)V_\mathrm{TTG}
    =
    \begin{pmatrix}
    h_{\sqrt{2}\alpha,\eta,\theta}(\vr)\\
    &   h_{\mathrm{D},1}(\vr)
    \end{pmatrix},
    \notag\\
    V_\mathrm{TTG}&=
    \frac{1}{\sqrt{2}}
    \begin{pmatrix}
    1   &   0   &   1   \\
    0   &   \sqrt{2}  &  0   \\
    1   &   0   &   -1 
    \end{pmatrix}.
\end{align}
First, we note that the TTG spectrum separated into independent sets of bands---a TBG-like 
set described by the two-layer Hamiltonian $h_{\sqrt{2}\alpha,\eta,\theta}$ 
(an $8\times8$ object when sublattice and spin are included) and an MLG-like Dirac cone 
described by $h_{\mathrm{D},1}$. Using the reasoning above, we expect a set of flat TBG-like 
bands to occur when $\sqrt{2}\alpha = 1/\sqrt{3}$. Equivalently, we can assign an 
effective TBG twist angle describing these bands, $\theta_\mathrm{TBG}^\mathrm{eff}=\theta/\sqrt{2}$ 
where $\theta$ is the physical twist angle of the system. 
If $\theta^\mathrm{eff}_\mathrm{TBG}$ is expected to yield flat 
bands in TBG, then we would similarly expect $\sqrt{2}\theta_\mathrm{TBG}^\mathrm{eff}$ 
to yield a set of flat bands in TTG.

{\theorysubsubsection{{Twisted quadrilayer graphene}}}
For four layers, we start with
\begin{align}
    h_\mathrm{TQG}(\vr)
    &=
    \begin{pmatrix}
    h_{\mathrm{D},1}(\vr)    &   T(\vr)  &   0   &   0   \\
    T^\dag(\vr) &   h_{\mathrm{D},2}(\vr)    &   T^\dag(\vr) &   0   \\
    0   &   T(\vr)  &   h_{\mathrm{D},1}(\vr)    &   T(\vr)  &   \\
    0   &   0   & T^\dag(\vr)   &   h_{\mathrm{D},2}(\vr)
    \end{pmatrix}.
\end{align}
With the appropriate change of basis, we obtain
\begin{align}\label{eqn:tqg_blk_diag}
    \tilde{h}_\mathrm{TQG}(\vr)
    &=
    V_\mathrm{TQG}^\dag h_\mathrm{TQG}(\vr)V_\mathrm{TQG}
    =
    \begin{pmatrix}
    h_{\varphi \alpha,\eta,\theta}(\vr)   \\
    &   h_{\varphi^{-1}\alpha,\eta,\theta}(\vr)
    \end{pmatrix},
    \notag\\
    V_\mathrm{TQG}&=
    \frac{1}{\sqrt{1+\varphi^2}}
    \begin{pmatrix}
    1 &   0   &   -\varphi    &   0   \\
    0   &   \varphi &   0   &   -1  \\
    \varphi &   0   &   1   &   0   \\
    0   &   1   &   0   &   \varphi
    \end{pmatrix},
\end{align}
where $\varphi=(1+\sqrt{5})/2$ is the golden ratio. 
In this case, we therefore expect the TQG spectrum to possess two sets of TBG-like bands 
characterized by effective TBG twist angles $\theta/\varphi$ and $\theta/\varphi^{-1}$.

{\theorysubsubsection{{Twisted pentalayer graphene}}}
The final system considered is the twisted pentalayer graphene. 
In the original layer basis, the Hamiltonian is
\begin{align}
    h_\mathrm{TPG}(\vr)
    &=
    \begin{pmatrix}
    h_{\mathrm{D},1}(\vr)    &   T(\vr)  &   0   &   0   &   0  \\
    T^\dag(\vr) &   h_{\mathrm{D},2}(\vr)   &   T^\dag(\vr) &   0   &   0   \\
    0   &   T(\vr)  &   h_{\mathrm{D},1}(\vr)    &   T(\vr)   &   0  \\
    0   &   0   & T^\dag(\vr)   &   h_{\mathrm{D},2}(\vr)   &   T^\dag(\vr)   \\
    0   &   0   &   0   &   T(\vr)  &   h_{\mathrm{D},1}(\vr)
    \end{pmatrix}.
\end{align}
Once more, independent, co-existing TBG- and MLG-like subsystems are revealed 
with the appropriate change of basis: 
\begin{align}
\label{eq:tpg_ham_sector_form}
    \tilde{h}_\mathrm{TPG}(\vr) 
    &=
    V^\dag_\mathrm{TPG}h_\mathrm{TPG}(\vr)V_\mathrm{TPG}
    =
    \begin{pmatrix}
    h_{\sqrt{3}\alpha,\eta,\theta}(\vr)     \\
    &   h_{\mathrm{D},1}(\vr)  \\
    &   &   h_{\alpha,\eta,\theta}(\vr)
    \end{pmatrix}
    \nt
    V_\mathrm{TPG}
    &=\frac{1}{\sqrt{6}}
    \begin{pmatrix}
    1   &   0   &   \sqrt{2}    &\sqrt{3}   &   0   \\
    0   &   \sqrt{3}    &    0   &   0  & \sqrt{3} \\
    2   &   0   &   -\sqrt{2}   &   0   &   0   \\
    0   &   \sqrt{3}    &   0   &   0   &   -\sqrt{3}   \\
    1&  0   &   \sqrt{2}    &   -\sqrt{3}   &   0
    \end{pmatrix}.
\end{align}
There are now \emph{two} independent TBG-like bands characterized by effective 
twist angles $\theta/\sqrt{3}$ and $\theta$ in addition to a MLG-like Dirac cone.

\theorysubsubsection{Model Parameters}
As indicated in Eq.~\eqref{eqn:vflat_approx}, the magic-angle value is essentially 
determined by the velocity of monolayer graphene $v_0$ and the interlayer tunnelling 
amplitude $w$. We fix $v_0$ for all considered configurations. The magnitude of the 
interlayer tunnelling amplitude is typically estimated to be around $\sim100$~meV. 
In case of TQG, a gap is expected to open at charge neutrality when finite $D$ field is 
applied. However, it onsets for any $|D|>0$ when the physical angle $\theta$ is less 
than $\varphi \theta_\mathrm{TBG}^\mathrm{magic}$, where $\theta_\mathrm{TBG}^\mathrm{magic}$ 
is the magic angle for TBG (as determined by $v_0$ and $w$). When $\theta$ is larger 
than $\varphi \theta_\mathrm{TBG}^\mathrm{magic}$, a gap still opens, but only above certain
finite $D$ fields. As \prettyref{exfig:CI_TTG} and \prettyref{exfig:gatemapT}j show, 
the latter scenario is observed in the TQG, device D2 (twist angle $1.8\degree$), leading 
us to select $w=108$~meV near the value used in Ref.~\citenum{bistritzerMoireBandsTwisted2011}. 
In particular, in the left panel of \prettyref{exfig:gatemapT}j, 
the $R_{xx}$ is plotted as a function of $D$ field, displaying non-monotonic 
behaviour---a resistance dip around $D/\epsilon_0\sim0.5$~V~nm$^{-1}$ followed by a steep 
increase at higher $D$, signalling the development of an insulating gap.
Analogous trends are repeated on the right of \prettyref{exfig:gatemapT}j, which 
shows an increase in $\nu=0$ DOS (corresponding to the resistance dip) followed 
by a decrease to zero DOS. 

Similar reasoning can be applied to the TTG and TPG samples, although it is slightly 
more nebulous since a non-interacting gap is not expected to open in TTG and TPG for 
any $D$ value at the CNP. Instead, when $\theta>\sqrt{2}\theta_\mathrm{TBG}^\mathrm{magic}$ 
for TTG and $\theta>\sqrt{3}\theta_\mathrm{TBG}^\mathrm{magic}$ for TPG, the system 
should become metallic with increasing $D$, whereas in the converse situation, the $D$ field should 
immediately gap out all states except for a dispersive MLG-like Dirac cone.
Following this line of reasoning, the results of \prettyref{exfig:gatemapT} suggest 
that the twist angle in TTG is \emph{below} the magic angle, whereas the one in the 
TPG sample is \emph{above} the magic angle. Accordingly, we select $w=110$~meV for TTG 
and $w=102$~meV for TPG modeling. Notably, the resistance behaviour and theoretical 
DOS shown in \prettyref{exfig:gatemapT}k for TPG are very similar to the results 
in \prettyref{exfig:gatemapT}j with the primary distinction being that the 
high-displacement field state does not display the activated transport of an 
insulator. Similarly, although not obvious from the DOS plot
itself, the band structure of TPG at large $D$ is semimetallic (as opposed to insulating).

The value of the interlayer AA hopping $w'$ is expected to be less than the interlayer AB 
hopping, as a result of lattice relaxation (see next section). We chose $w'=60$~meV for 
all three multilayers considered, which is in agreement with the estimates 
of Ref.~\citenum{ledwithTBNotTB2021} and similar to values used previously for 
TBG/WSe$_2$ structures\cite{aroraSuperconductivityMetallicTwisted2020}. 

We note that, while consistent with experiment in the fashion outlined about, 
other factors could be also at play, modifying the behaviour at CNP in ways 
not captured by our analysis. Ultimately, however, the choices made here are 
not expected to greatly influence any of the results in this section. 

\theorysubsection{Relative stacking}
\label{methods:stacking}
An important distinction between TBG and graphene moir\'e heterostructures
containing additional layers is the band structure dependence of the relative
layer displacement.
Not only must the graphene sheets be stacked with alternating angles, as
discussed in the main text and in the previous section, but moreover, 
the emergence of independent TBG- and Dirac-like bands only occurs when all odd 
(even) layers are AA stacked, \emph{i.e.}, stacked directly on top of one another.
As stated above, we envision grouping the layers into odd and even sets, each 
stacked rigidly atop one another. The two sets are then twisted relative to one 
another by the twist angle $\theta$. 
We have assumed that this stacking was realized in the previous presentation 
and now argue for the feasibility of this assumption.

In TTG, it has been numerically shown that this situation is energetically 
preferable: the system naturally relaxes into the odd/even aligned stacking 
configuration\cite{carrUltraheavyUltrarelativisticDirac2020}.
This result is further experimentally verified in 
transport\cite{parkTunableStronglyCoupled2021} and local 
probe\cite{kimSpectroscopicSignaturesStrong2021} measurements. A simple
heuristic supports these results and permits a generalization to additional
layers. Starting from a bilayer system, the moir\' e superlattice is manifest on
the microscopic lattice scale as the periodic variation of the relative
interlayer stacking: one has AA regions at the moir\' e hexagon centres, while AB
and BA stacking regions represent the moir\' e hexagon vertices. The AA regions have
a relatively high energy compared to the Bernal-like region and the lattice
accordingly responds by relaxing to minimize their area. We now consider adding
a third layer with the same relative twist angle as the first layer, but for the
moment arbitrarily displaced from that layer. A moir\' e superlattice is of course
also generated between the new layer and the second layer, and the system once
again seeks to minimize (maximize) the area of the AA (AB/BA) regions.
Crucially, if the first and third layers are misaligned, the AA regions between
the first and second layers are misaligned from the AA regions between the
second and third layers, frustrating the ability of the lattice to relax. Only
when the first and third layers are aligned will the AA region occur at the same
locations and only then can the system optimize its energy through relaxation.
These arguments clearly generalize to quadrilayer and pentalayer systems---we
need only consider the moir\' e pattern generated by each adjacent pair of graphene
sheets to conclude that relaxation is optimized by an odd/even aligned
configuration. (A complementary explanation is that TTG is necessarily an
intermediate step in the construction of the TQG and TPG devices, and thus the
odd alignment is already baked into a subset of the layers.)

\theorysubsection{Mirror symmetry}
\label{methods:mirrorsymmetry}

In the systems with an odd number of layers, an onsite mirror symmetry is 
present, which acts as

\begin{align}
    \psi(\vr)&\to U_\mathrm{mirror}^{(n_\mathrm{layer})}\psi(\vr),
    &
    n_\mathrm{layer}\text{ odd},
\end{align}
where
\begin{align}\label{eqn:mirror_operator}
    \left[U_{\mathrm{mirror}}^{(n_\mathrm{layer})}\right]_{\ell,\ell'}
    &=
    \delta_{\ell,n_\mathrm{layer}-\ell'+1}.
\end{align}
Here, $\ell$, $\ell'$ label the system's layers.
Effectively, this operator simply flips the layers around, for instance, 
interchanging layers 1 and 3 in TTG, while keeping the middle layer fixed.
In terms of the matrices, this invariance manifests simply as the relation 
$[h_\mathrm{TTG}(\vr),U_\mathrm{mirror}^{(3)}]=0$ and $[h_\mathrm{TPG}(\vr),U_\mathrm{mirror}^{(5)}]=0$.
As we see below, the preservation of this symmetry is inextricably tied to the 
block diagonal form of the TTG continuum model presented in Eq.~\eqref{eqn:ttg_blk_diag}.
In particular, rotating $U_\mathrm{mirror}^{(3)}$ to the subsystem basis defined 
by $V_\mathrm{TTG}$ returns 
$\tilde{U}_\mathrm{mirror}^{(3)}=V_\mathrm{TTG}^TU_\mathrm{mirror}^{(3)}V_\mathrm{TTG}=\mathrm{diag}(1,1,-1)$.
The TBG-like subsystem corresponds precisely to the \emph{even parity} sector 
(\emph{i.e.}, has eigenvalue $+1$ under the action the mirror symmetry) whereas the 
dispersive MLG-like subsystem belongs to the \emph{odd} parity sector 
(\emph{i.e.}, has eigenvalue $-1$ under the action the mirror symmetry).
The TBG- and MLG-like bands thus cannot hybridize without breaking this symmetry.

We can similarly rotate the TPG operator, $U^{(5)}_\mathrm{mirror}$ to the subsystem 
basis $\tilde{U}_\mathrm{mirror}^{(5)}=V_\mathrm{TPG}^TU_\mathrm{mirror}^{(5)}V_\mathrm{TPG}$, 
yielding $\tilde{U}_\mathrm{mirror}^{(5)}=\mathrm{diag}(1,1,1,-1,-1)$.
Comparing against Eq.~\eqref{eq:tpg_ham_sector_form}, we observe that \emph{both} the 
TBG-like subsystem with effective twist angle $\theta/\sqrt{3}$ and the MLG-like 
subsystem belong to the even parity sector, whereas the subsystem with effective 
twist angle $\theta$ belongs to the odd parity sector. The mirror symmetry therefore 
only protects the latter subsystem---which is notably \emph{not} at the magic angle in the experiment.
In other words, in TPG with mirror symmetry, flat TBG-like band and MLG-like band can hybridize (while 
the dispersive TBG-like band is protected).

A mirror-like symmetry also exists for even-layered systems like TBG and TQG, but it does 
not act in an onsite fashion and is therefore not useful for the analysis that follows.
For TQG, hybridization between subsystems is therefore not prohibited by symmetry.

\theorysubsection{Band mixing}
\label{methods:band_mixing}
In obtaining the independent TBG- and MLG-like bands (subsystems) listed above, a 
number of assumptions were made and one may be concerned about the relative 
robustness of these results. For TTG, at least, this question may be dismissed 
so long as mirror symmetry is present; above, we showed that this mirror symmetry 
protects the block diagonal subsystem form obtained for TTG.
Similarly, mirror symmetry disallows mixing in TPG between certain (but not all) 
subsystems. However, the mirror symmetry is explicitly broken by the application 
of a displacement field as well as by the WSe$_2$ substrate used in the experiment. 
Below, we show that these modifications induce mixing between all subsystems.
We additionally consider other mirror-preserving effects that may result in 
subsystem mixing in TQG and TPG. 

Besides the displacement field, we find that the subsystem-mixing energy scales 
discussed below are relatively small compared to the input parameters of the 
continuum model, i.e., compared to an interlayer tunnelling of $w$ and $w'$. 
More importantly, they are also smaller than the observed bandwidth of TBG, which 
spectroscopic measurements indicate is $\sim40$~meV for samples close to the magic 
angle\cite{kerelskyMaximizedElectronInteractions2019,choiElectronicCorrelationsTwisted2019,
xieSpectroscopicSignaturesManybody2019,jiangChargeOrderBroken2019}.
The subleading magnitude of the effects we explore below thus bolsters our use of 
the alternating-angle continuum model, at least as a starting point. We note that 
the relatively small subsystem hybridization discussed here could be significantly 
magnified by interactions.

\theorysubsubsection{Effect of displacement field}
In the main text, we allude to the fact that a finite displacement field 
mixes the TBG- and MLG-like subsystems obtained in the previous sections.
This effect is included in the Hamiltonian through the addition of

\begin{align}\label{eqn:disp_ham_def}
    H_\mathrm{disp}
    &=
    \sum_{\ell,\ell'=1}^{n_\mathrm{layer}} \int d^2\vr\, 
    \psi^\dag_\ell(\vr)\left[h^{(n_\mathrm{layer})}_\mathrm{disp}\right]_{\ell,\ell'}\psi_{\ell'}^\dag(\vr),
    &
    \left[h^{(n_\mathrm{layer})}_\mathrm{disp}\right]_{\ell,\ell'}
    &=
    U\delta_{\ell,\ell'}\left( \frac{1}{2} - \frac{\ell-1}{n_\mathrm{layer}-1} \right).
\end{align}

Specifically, we have $h^{(3)}_\mathrm{disp}=(U/2)\mathrm{diag}(1,0,-1)$, $h_\mathrm{disp}^{(4)}=U\mathrm{diag}(1/2,1/6,-1/6,-1/2)$, and $h^{(5)}_\mathrm{disp}=U\mathrm{diag}(1/2,1/4,0,-1/4,-1/2)$.
Here, the scale $U$ is defined as outlined in section~\ref{methods:ttg_tqg_gap}. 

Focusing first on the odd-layered systems, TTG and TPG, we observe that this 
perturbation explicitly breaks the mirror symmetry introduced in the previous section.
In particular, $h^{(3)}_\mathrm{disp}$ and $h^{(5)}_\mathrm{disp}$ \emph{anticommute} with 
$U^{(3)}_\mathrm{mirror}$ and $U^{(5)}_\mathrm{mirror}$ respectively: 
$\{h_\mathrm{disp}^{(3)},U_\mathrm{mirror}^{(3)}\}=0$ and $\{h_\mathrm{disp}^{(5)},U_\mathrm{mirror}^{(5)}\}=0$.
The displacement field therefore allows subsystems within different parity sectors to hybridize.
The effect of this addition is apparent when $h_\mathrm{disp}^{(n_\mathrm{layer})}$ is tranformed to the subsystem basis of Eqs.~\eqref{eqn:ttg_blk_diag} and~\eqref{eq:tpg_ham_sector_form}:

\begin{align}
    \tilde{h}_\mathrm{disp}^{(3)}
    &=V_\mathrm{TTG}^\dag h_\mathrm{disp}^{(3)}V_\mathrm{TTG}    
    =\frac{U}{2}\begin{pmatrix}
    0   &   0   &   1 \\
    0   &   0   &   0   \\
    1 &   0   &   0
    \end{pmatrix}
    \begin{matrix}
        \left.\begin{matrix}
         \phantom{\!}   \\ \phantom{\!}
        \end{matrix}\right\}
        &
        \!\!\text{Parity-even sector}
        \\
        \left.\phantom{\!} \right\}  
        &   
        \!\!\text{Parity-odd sector}
    \end{matrix}
    \nt
    \tilde{h}_\mathrm{disp}^{(5)}
    &=V_\mathrm{TPG}^\dag h_\mathrm{disp}^{(5)}V_\mathrm{TPG}    
    =\frac{U}{4\sqrt{3}}
    \begin{pmatrix}
    0   &   0   &   0   &   2  &    0   \\
    0   &   0   &   0   &   0   &   \sqrt{3}    \\
    0   &   0   &   0   &   2\sqrt{2}   &   0   \\
    2   &   0   &   2\sqrt{2}   &   0   &   0   \\
    0   &   \sqrt{3}    &   0   &   0   &   0
    \end{pmatrix}
    \begin{matrix}
        \left.
        \begin{matrix}
         \phantom{\!}   \\ \phantom{\!} \\ \phantom{\!}
        \end{matrix}\right\}
        &
        \!\!\text{Parity-even sector}
        \\
        \left.
        \begin{matrix}
        \phantom{\!}    \\  \phantom{\!}
        \end{matrix}\right\}
        &   
        \!\!\text{Parity-odd sector}
    \end{matrix}
\end{align}

We thus explicitly see the way in which the displacement field induces mixing between subsystems.

Subsystem mixing is also a natural consequence of the displacement field in TQG. 
The Hamiltonian in Eq.~\eqref{eqn:disp_ham_def} takes the form $h_\mathrm{disp}^{(4)}=U\mathrm{diag}(1/4,1/6,-1/6,1/4)$, which becomes

    \begin{align}
        \tilde{h}_\mathrm{disp}^{(4)}
        &=
        \frac{U}{30\varphi}(1+\varphi^2)
        \begin{pmatrix}
        \varphi^{-3}    &   0   &   -4  &   0   \\
        0   &   -\varphi^{-3}   &   0   &   -4  \\
        -4  &   0   &   \varphi^3   &   0   \\
        0   &   -4  &   0   &   -\varphi^3
        \end{pmatrix}
        \begin{matrix*}[l]
            \left.\
            \begin{matrix}
            \phantom{\!}   \\ \phantom{\!}
            \end{matrix}
            \right\}
            &
            \!\!\text{Subsystem with $\theta^\mathrm{eff}_\mathrm{TBG}=\varphi\theta$}
            \\           
            \left.\
            \begin{matrix}
            \phantom{\!}   \\ \phantom{\!}
            \end{matrix}
            \right\}
            &   
            \!\!\text{Subsystem with $\theta^\mathrm{eff}_\mathrm{TBG}=\varphi^{-1}\theta$}
        \end{matrix*}
    \end{align}
in the subsystem basis.

\theorysubsubsection{Proximity-induced spin-orbit coupling}
One of the exterior layers of the samples considered here is placed adjacent to WSe$_2$.
This type of construction was first shown to induce spin-orbit coupling in twisted bilayer graphene in 
Ref.~\citenum{aroraSuperconductivityMetallicTwisted2020}.
The presence of WSe$_2$ breaks not only the spin symmetry, but also the 
mirror symmetry in systems considered here and possibly induces subsystem mixing.
The magnitude of the induced spin-orbit scale has been measured to be 
approximately $1-5$~meV in TBG, smaller than the other scales of the theory (e.g., the band width).
In effect, in rotating to the subsystem basis, the spin-orbit terms 
are ``spread'' across an increasing number of layers by the unitary 
transformations $V_\mathrm{TTG}$, $V_\mathrm{TQG}$, $V_\mathrm{TPG}$.

\theorysubsubsection{Mirror-symmetric, nonuniform charge distribution}: 
\label{methods:mirror_symmetry_chem_variation}
The chemical potentials of the different layers may also differ in a 
way that is symmetric under onsite mirror actions $U_\mathrm{mirror}^{(4/5)}$ of Eq.~\eqref{eqn:mirror_operator}. 
In particular, we may have $h_{\mu\text{-}\mathrm{var}}^{(4)}=\mathrm{diag}(\delta\mu_1,\delta\mu_2,\delta\mu_2,\delta\mu_1)$, which takes the subsystem-basis form

\begin{align}
    \tilde{h}_{\mu\text{-}\mathrm{var}}^{(4)}
    &=
    \frac{\delta\mu}{\sqrt{5}}
    \begin{pmatrix}
    -1  &   0   &   -2  &   0   \\
    0   &   -1  &   0   &   2   \\
    -2  &   0   &   1   &   0   \\
    0   &   2   &   0   &   1
    \end{pmatrix},
\end{align}
where $\delta\mu=\delta\mu_1=-\delta\mu_2$.
Similarly, for TPG, a term like 
$h_\mathrm{\mu\text{-}\mathrm{var}}^{(5)}=\mathrm{diag}(\delta\mu_1,\delta\mu_2,\delta\mu_3,\delta\mu_2,\delta\mu_1)$ also preserves the mirror operator $U_\mathrm{mirror}^{(5)}$ but 
can be shown to induce inter-subsystem mixing within the even parity sector.
As we demonstrate in section~\ref{method:interactions_in_tpg}, such a term 
is naturally generated by the Coulomb interaction.
We specify to TPG in that section, but the reasoning is analogous for TQG 
(and for TTG, although this term will not induce mixing between sectors because 
of the mirror symmetry).

Although generically present, the Coulomb interaction-generated terms of this 
form are relatively small compared to the other terms present. 
The calculations presented below estimate that values of $|\delta\mu_\ell|<5-10$~meV 
for TPG are generated as one dopes away from charge neutrality.
We expect the results for TQG to follow the same trend.

\theorysubsubsection{Next-nearest layer tunnelling} 
Our Hamiltonian so far only includes tunnelling between neighbouring layers. 
Generically, however, hopping between next-nearest neighbouring layers occurs as well. 
For TQG, we could therefore consider hopping between layers 1 (2) and 4 (3):

\begin{align}
    h_\mathrm{nnl}^{(4)}
     &=
    \begin{pmatrix}
        0   &   0   &   T_\mathrm{nnl}    &   0   \\
        0   &   0   &   0   &   T_\mathrm{nnl} \\
        T^\dag_\mathrm{nnl}    &   0   &   0   &   0   \\
        0   &   T^\dag_\mathrm{nnl}    &   0   &   0   
    \end{pmatrix}.
\end{align}

Assuming for simplicity that
$T_\mathrm{nnl.}=T^\dag_\mathrm{nnl}$, in the subsystem basis, this term takes the form
\begin{align}
    \tilde{h}_\mathrm{nnl}^{(4)}
    &=
    \frac{1}{\sqrt{5}}
    \begin{pmatrix}
    2T_{\mathrm{nnl}}   &   0   &   - T_\mathrm{nnl}    &   0   \\
    0   &   2T_\mathrm{nnl}   &   0   &    T_\mathrm{nnl}  \\
    -T_\mathrm{nnl}    &   0   &   -2T_\mathrm{nnl} &   \\
    0   &   T_\mathrm{nnl}   &   0   &   -2T_\mathrm{nnl}
    \end{pmatrix}.
\end{align}

Note that because we assume next-nearest layers are stacked AA relative to one another, 
to first order, no spatial dependence is expected in $T_\mathrm{nnl}$.
The subsystems are similarly mixid with the five-layer analogue $h^{(5)}_\mathrm{nnl}$.
Reference~\citenum{carrUltraheavyUltrarelativisticDirac2020} computed the values 
of $T_\mathrm{nnl}$ expected in TTG (where it will not induce subsystem mixing) and 
found that a typical scale $|T_{\mathrm{nnl},ij}|\sim 5-10$~meV, which translates to 
$|T_{\mathrm{nnl},ij}|/\sqrt{5}\sim 3-5$~meV ($i$ and $j$ are sublattice indices).

\theorysubsubsection{Lattice relaxation}
As mentioned in section~\ref{methods:stacking}, the moiré lattice relaxes in order to 
minimize AA regions and maximize AB/BA regions. 
As mentioned, this relaxation effect ultimately depresses the value of $w'$ 
(interlayer AA tunnelling) relative to $w$ (interlayer AB/BA tunnelling) as a 
result of out-of-plane corrugation.
For interior layers, which neighbour more than a single sheet, the effects of 
relaxation are naturally stronger than for exterior layers. Consequently, 
the value of $\eta=w'/w$ appropriate for tunnelling to and from interior layers is reduced.
Our assumption below Eq.~\eqref{eq:total_hamiltonian} that $T_{\ell,\ell+1}(\vr)$ 
depended only on whether $\ell$ was even or odd is no longer valid.
Unsurprisingly, this effect once again mixes the subsystems in TQG and TPG.
Reference~\citenum{ledwithTBNotTB2021} estimated the magnitude of this 
effect and determined that it should be in the range $5-10$~meV for the 
twist angles considered here.

\theorysubsection{The role of interactions in TPG}
\label{method:interactions_in_tpg}
The presence of flat-band subsystem in the low-energy theory of the multilayer
graphene structures necessitates the consideration of interaction-driven band
structure corrections. In the following, we focus specifically on the case of TPG
as its phase diagram demonstrates the strongest deviation from the minimal paradigm that
a multilayer structure maybe thought of as a TBG-like Hamiltonian with spectating
additional bands. Rather, as we argue, the dispersive TBG- and MLG-like subsystems 
play a crucial role in extending the filling range of the superconducting pocket 
in accordance with scenario ($ii$) and ($iii$). Here, we consider three types
of interaction corrections: (a) an in-plane Hartree correction; 
(b) a two-parameter effective model mimicking generic Hartree-Fock modifications of
band structure; (c) an out-of-plane Hartree correction allowing for inhomogeneous
charge distribution between the layers. We demonstrate that these
effects generically lead to two consequences for the electronic spectrum: 
promoting charge redistribution to the non-flat bands and also leading to possible
symmetry breaking between the non-flat and flat bands.

\label{methods:Hartree} \theorysubsubsection{Hartree correction} We begin with an in-plane
Hartree effect. As demonstrated experimentally in previous work on 
TBG\cite{choiInteractiondrivenBandFlattening2021} and 
TTG\cite{kimSpectroscopicSignaturesStrong2021}, filling-dependent
interaction effects, specifically Hartree and Fock corrections, drastically
alter the electronic dispersion. Here we incorporate only a Hartree
mechanism\cite{guineaElectrostaticEffectsBand2018,rademakerChargeSmootheningBand2019,goodwinHartreeTheoryCalculations2020,
  calderonInteractions8orbitalModel2020} in the analysis, arguing that its key
effect, a relative shift of flat bands up in energy with respect to the non-flat
bands, is the simplest mechanism through which the size of the superconducting
pocket in TPG is extended. We then supplement this discussion with a phenomenological
Hartree-Fock-like theory. Before proceeding, we stress that the main purpose of the 
analysis in this section is to demonstrate that scenario ($i$) wherein flat bands 
are filled only to $\nu_\mathrm{flat}\approx +3$ at $\nu\approx +5$ is highly unlikely, thus highlighting the non-trivial role played by the dispersive TBG- and MLG-like bands.

The foundations of the Hartree calculation in TPG described below are identical to the analysis in
Refs.~\citenum{choiInteractiondrivenBandFlattening2021} and~\citenum{kimSpectroscopicSignaturesStrong2021}. We reproduce them here for the convenience of the reader. We introduce the Coulomb
interaction into the system through

\begin{align}\label{eqn:CoulombInt}
    H_C&=
    \frac{1}{2}\int d^2\vr \,d^2\vr' \,
    \delta\rho(\vr)V(\vr-\vr')\delta\rho(\vr').
\end{align}

In section~\ref{methods: continuum_model}, we introduced creation and annihilation 
operators, $\psi^\dag(\vr)$ and $\psi(\vr)$, that correspond to the non-interacting 
eigenstates given by the Hamiltonian of Eq.~\eqref{eq:total_hamiltonian}. 
Here and in what follows, we suppress the layer, valley, sublattice and spin subscripts.
In Eq.~\eqref{eqn:CoulombInt}, $V(\vr)=e^2/(\epsilon |\vr|)$ is the Coulomb potential and
$\delta\rho(\vr)=\psi^\dag(\vr)\psi(\vr)-\rho_\mathrm{CN}(\vr)$, where
$\rho_\mathrm{CN}(\vr)=\langle \psi^\dag(\vr)\psi(\vr)\rangle_\mathrm{CN}$ is
the expectation value of the density at the charge-neutrality point.  The use of
$\delta\rho(\vr)$ instead of $\rho(\vr)$ in the interaction is motivated by the
expectation that the input parameters of the model $H_{\mathrm{T}n_\mathrm{layer}G}=H_\mathrm{cont}$ already
include the effect of interactions at the charge-neutrality point. The
dielectric constant $\epsilon$ in the definition of $V(\vr)$ is used as a
fitting parameter; see discussion below for details. 

We study the effect of the interacting continuum model of magic-angle TPG through a
self-consistent Hartree mean-field calculation. Instead of solving the many-body
problem, we obtain the quadratic Hamiltonian that best approximates the full
model when only the symmetric contributions of $H_C$ are included, i.e., the
Fock term is neglected. Thus instead of
$H_\mathrm{cont}+H_C$, we study the Hamiltonian

\begin{align}\label{eqn:Hmf_hartree}
    H_\mathrm{MF}^{(\nu)}
    &=
    H_\mathrm{cont}
    + 
    H^{(\nu)}_\mathrm{H}
    -
    \frac{1}{2}\langle H_\mathrm{H}^{(\nu)} \rangle_\nu,
\end{align}
where $H_\mathrm{H}^{(\nu)}$ is the Hartree term at filling $\nu$,
\begin{align}
\label{eqn:hartree_term}
    H_\mathrm{H}^{(\nu)}
    &=
    \int_{\vk,\vk',\vq}V(\vq)
    \langle \psi^\dag(\vk'+\vq)\psi(\vk') \rangle_{\nu}
    \psi^\dag(\vk)\psi(\vk-\vq),
\end{align}
and the last term in Eq.~\eqref{eqn:Hmf_hartree} simply ensures there is no
double counting when one calculates the total energy. In the above equation,
$V(\vec{q})=2\pi e^2/(\epsilon |\vq|)$ is the Fourier transform of the Coulomb
interaction $V(\vr)$ in Eq.~\eqref{eqn:CoulombInt}, and the expectation value
$\langle\hat{\mathcal{O}}\rangle_\nu = \langle\hat{\mathcal{O}}\rangle_\mathrm{occ} - \langle\hat{\mathcal{O}}\rangle_\mathrm{CN}$
only includes states that are filled up to $\nu$ \emph{relative} to charge
neutrality, {as defined by diagonalizing the Hamiltonian
  $H_\mathrm{MF}^{(\nu)}$}. 
  
Typically, for a ``jellium''-like model, the expectation value in
Eq.~\eqref{eqn:hartree_term} vanishes save for $\vq=0$, which is subsequently
cancelled by the background charge---allowing one to set $V(\vq=0)=0$ and
completely ignore the Hartree interaction. However, because the moir\'e pattern
breaks continuous translation symmetry, momentum is only conserved modulo a
reciprocal lattice vector. We therefore obtain

\begin{align}\label{eqn:Hartree1}
    H_\mathrm{H}^{(\nu)}
    &=
    \sum_{\vec{G}}'
    V(\vec{G})
    \int_{\vk'}
    \langle \psi^\dag(\vk'+\vec{G})\psi(\vk') \rangle_{\nu}
    \int_{\vk}
    \psi^\dag(\vk)\psi(\vk-\vec{G}),
\end{align}
where the prime above the summation over the moir\'e reciprocal lattice vectors
indicates that $\vec{G}=0$ is excluded. The self-consistent procedure begins by
assuming some initial value of $H_\mathrm{H}^{(\nu)}$ and diagonalizing the
corresponding mean-field Hamiltonian $H_\mathrm{MF}^{(\nu)}$ to obtain the Bloch
wavefunctions and energy eigenvalues. These quantities are then used to re-compute
the expectation values that define $H_\mathrm{H}^{(\nu)}$ and thus
$H_\mathrm{MF}^{(\nu)}$ subject to the cascade treatment described above. This
process is repeated until one obtains the quadratic Hamiltonian
$H_\mathrm{MF}^{(\nu)}$ that yields the correlation functions
$\langle\cdot\rangle_\nu$ used in its definition.

Due to the accumulation of electronic density at the $AA$ sites of the
stacking sequence, the Hartree potential is dominated by the first `star' of moir\'e reciprocal
lattice vectors\cite{guineaElectrostaticEffectsBand2018,ceaElectronicBandStructure2019}, which in our conventions corresponds to
$\vec{G}_n=R\big(2\pi(n-1)/6\big)\frac{4\pi}{\sqrt{3}L_M}(1,0)^T$ for
$n=1,\dots,6$, with $R(\phi)$ a rotation matrix. 
The restriction to the $\vec{G}_n$'s paired with the $2\pi/6$ rotation symmetry of the continuum model greatly
simplifies the calculation of the Hartree term.
Notably,
$V(\vec{G})\int_{\vk'}\langle\psi^\dag(\vk'+\vec{G})\psi(\vk')\rangle_\nu$ must
be the same for all $\vec{G}_n$, and, instead of Eq.~\eqref{eqn:Hartree1}, we
use

\begin{align}
\label{eq:sc_equations}
    H_\mathrm{H}^{(\nu)}
    &=
    V_\mathrm{H}^{(\nu)}
    \sum_{n=1}^6 \int_\vk \psi^\dag(\vk)\psi(\vk-\vec{G}_n),
    &
    V_\mathrm{H}^{(\nu)}
    &=
    \frac{1}{6}\sum_{n=1}^6V(\vec{G}_n)\int_{\vk'}\langle\psi^\dag(\vk'+\vec{G})\psi(\vk')\rangle_\nu\,.
\end{align}

The self-consistent procedure in this case is identical to that described in the
previous paragraph, but due to the reduced number of reciprocal lattice vectors
that are included in the summation, the calculation is computationally easier.
Convergence is typically reached within $\sim 6$ iterations.

We now proceed to discuss the precise effect of the Hartree correction. Since the Hartree
correction couples bare graphene states at momenta $\vec{k}$ and $\vec{k}+\vec{G}$, its
effect is most pronounced for subsystems of the Hamiltonian whose eigenstates require
multiple bare graphene states originating from multiple moir\'e BZs, e.g. $\vec{k}+\vec{G}$
states with $\vec{G}$ extending beyond the second BZ. As such, Hartree affects the flat-band
subsystem most severely since its eigenstates deviate the most from the bare
graphene states, while the MLG-like subsystem is affected the least. As a result, this
correction gives rise to an energy offset that shifts the flat bands upwards in
energy with respect to the rest of the energy spectrum (technically the
dispersive TBG-like subsystem is also shifted slightly with respect to the MLG-like subsystem). This
effect has been seen both theoretically and experimentally in
TTG\cite{fischerUnconventionalSuperconductivityMagicAngle2021,kimSpectroscopicSignaturesStrong2021}.
Thus we expect it to be present in TPG, as is confirmed through our
simulations; see \prettyref{exfig:theory_figure_interactions}a,b. Physically, this effect arises simply
because the charge distribution from the non-flat subsystems is more homogeneous in the unit cell
and, therefore, it contributes less to the potential of Eq.~\eqref{eqn:hartree_term}.

We now discuss what happens when one starts from charge neutrality and electron dopes the system.
Due to the shift of the flat band upwards in energy relative to the non-flat bands, more
charge can enter the non-flat bands upon doping (increasing $\nu$) than a na\"ive non-interacting model
predicts. 
As a result, the filling range of the flat TBG-like band
superconducting pocket may be extended since the filling of the flat bands $\nu_\mathrm{flat}$ can continue to lie in the range amenable to superconductivity,
whilst the total filling $\nu$ increases by adding charge to the non-flat bands.
This
is the central idea behind the scenarios ($i$) and ($ii$) discussed in section \ref{methods:scenario}. 

In the simulations for flat-band filling $\nu_\mathrm{flat} > +2$, we consider two 
ways to fill the otherwise 4-fold degenerate bands: an uncascaded model where all 
4-fold degenerate bands are filled equally, and a simple cascade model where two of 
the flat-band flavours (say spin $\uparrow$ for $K$, $K'$) are shifted down in energy 
such that the highest energy of the shifted bands falls on the Dirac point of the 
unshifted flat bands, c.f. \prettyref{exfig:theory_figure_interactions}c. In the absence 
of Hartree-induced band inversion (as in fact we will consider in the following 
section), the shifted bands ($\uparrow$ bands) are fully filled at $\nu_\mathrm{flat}=+2$ 
and the unshifted bands ($\downarrow$) and the dispersive bands contain the 
remaining $\nu-2$ charge. With the Hartree-induced gamma point inversion, 
the two sets of the flat bands (shifted - $\uparrow$, unshifted - $\downarrow$) 
become partially filled near $\nu_\mathrm{flat}=+2$. The shifted ($\uparrow$) 
band is mostly filled and the unshifted ($\downarrow$) is mostly empty. 
This simple approach qualitatively reproduces the effect of a
cascade at $|\nu_\mathrm{flat}|\approx2$ under the assumption that the 
specific nature of the cascade state (i.e. spin or valley polarized) is irrelevant for the
consideration of the total filling. We caution, however, that a cascade 
transition is an effect originating from an interplay between Hartree and Fock corrections 
(see Ref.~\citenum{choiInteractiondrivenBandFlattening2021} for further discussion), 
and that Fock corrections, which we neglected so far, can give rise to many 
effects.

The most crucial ones, including bandwidth broadening
\cite{ceaElectronicBandStructure2019,ceaBandStructureInsulating2020, 
xieWeakfieldHallResistivity2020, liuNematicTopologicalSemimetal2021}
and gap opening in the flat-band subsystem\cite{choiElectronicCorrelationsTwisted2019,
xieNatureCorrelatedInsulator2020,ceaBandStructureInsulating2020, bultinckGroundStateHidden2020,
christosCorrelatedInsulatorsSemimetals2021, lianTwistedBilayerGraphene2021, xieTwistedSymmetricTrilayer2021, kangStrongCouplingPhases2019}, 
may actually affect the charge distribution across the 
different subsystems. These effects are neglected in the current analysis---an approximation 
we will justify in the following section.

The cascade, as shown in \prettyref{exfig:theory_figure_interactions}g, allows
charge to enter the unfilled flat bands more easily compared to the uncascaded
ground state. This behaviour is expected since a cascade minimizes the contribution of
the Hartree term by redistributing charge away from parts of the flat bands
which overlap more strongly with the Hartree potential---in particular, for the
parameters considered here and within the relevant range of filling factors, 
the cascade is the ground state solution.
Note, however, that while including only the Hartree correction is sufficient to 
initiate cascade, Fock must be included in order for it to persist over the 
experimentally observed range (see Ref.~\citenum{choiInteractiondrivenBandFlattening2021} 
for further discussion on the interplay of Hartree and Fock corrections and the onset of cascade). 

To quantitatively estimate the fillings of the different subsystems,
it is necessary to parametrize the strength of the Coulomb interaction, e.g., the
dielectric constant $\epsilon$ that enters into Eq.~\eqref{eqn:hartree_term}. Although, in principle, the dielectric constant is fixed primarily by the substrate and any interaction corrections can be
accounted for via a self-consistent treatment, in practice\cite{guineaElectrostaticEffectsBand2018,
  ceaElectronicBandStructure2019,xieWeakfieldHallResistivity2020,choiInteractiondrivenBandFlattening2021}, it can be treated as a fitting parameter. If a bare value of the
interaction is used, then the resulting interaction corrections are too large
and lead to unobserved predictions\cite{choiInteractiondrivenBandFlattening2021,kimSpectroscopicSignaturesStrong2021}.
We use the cascade near $\nu_\mathrm{flat}\approx+2$ to constrain $\epsilon$.
We identify $\nu_\mathrm{flat}=+2$ with the experimental onset of the cascade transition, which occurs near $\nu\approx +2.15$; see \prettyref{fig:Fig4}e.
By choosing $\epsilon$ so that $\nu_\mathrm{flat}$ reaches $+2$ at the same point the total filling $\nu$ reaches $+2.15$ (see \prettyref{exfig:theory_figure_interactions}e), we find $\epsilon\approx 11.15$.
We emphasize that although this
is an approximate fitting relying on the particular model of a cascade, the Hartree-induced flat-band energy shift is a robust and important effect (\prettyref{exfig:theory_figure_interactions}f). Using the value of $\epsilon\approx 11.15$, we find that at $\nu\approx+5$, the flat bands 
are filled to approximate $\nu_\mathrm{flat}\approx +3.8$ (see \prettyref{exfig:theory_figure_interactions}g), further demonstrating
the implausibility of scenario ($i$).

We note that due to the hybridization of different bands under
a finite displacement field, the assignment of flat TBG-, 
dispersive TBG- and MLG-like subsystem becomes, to some degree, arbitrary as bands start 
to hybridize. For the purpose
of qualitative discussion, however, we can evaluate the spectral overlap of each finite-field
eigenstate with the zero-field basis and assign a label of ``flat/dispersive TBG-like/MLG-like"
based on the largest overlap. Within this convention, we find
that the displacement field enables easier charge accumulation in
the ``flat'' subsystem as opposed to the ``non-flat'' subsystems, thus 
suppressing the total filling range over which superconductivity can reach (see \prettyref{exfig:theory_figure_interactions}h).

\label{methods:hartree_and_fock_corrections}
\theorysubsubsection{Constraining Hartree and Fock} Preferential filling of the 
dispersive TBG- and MLG-like subsystems is also enabled by other interaction effects, 
for example, gap opening due to the Fock correction. This term, as mentioned previously, 
also plays a key part in the symmetry-breaking cascade as well as band broadening. 
While a careful microscopic analysis of Hartree and Fock effects in multilayer 
devices is necessary, here we introduce a simple phenomenological model 
intended to capture the qualitative effects of Hartree and Fock corrections 
on the filling of the non-flat subsystems. We hope that the simple 
parametrization of this model can be used as a benchmark for its validity against a more rigorous analysis.

To mimic the effects of Hartree and Fock, we add two additional ingredients 
to the non-interacting model of Eq.~\eqref{eq:tpg_ham_sector_form}: a constant 
energy shift of the flat-band subsystem $\Delta_H$ and an intralayer 
sublattice potential $\Delta_F \sigma^z$
\begin{align}
    \tilde{h}_\mathrm{TPG}(\vr) 
    &=
    \begin{pmatrix}
    \Delta_H \mathbb{1}_{2\times 2}+ \Delta_F \sigma^z + h_{\mathrm{D},1}(\vr)    &   \sqrt{3}T(\vr)  \\
    \sqrt{3}T^\dag(\vr) &   \Delta_H \mathbb{1}_{2\times 2}+ \Delta_F \sigma^z+h_{\mathrm{D},2}(\vr)    \\
    &   &   h_{\mathrm{D},1}(\vr)    \\
    &   &   &   h_{\mathrm{D},1}(\vr)    &   T(\vr)  \\
    &   &   &   T^\dag(\vr) &   h_{\mathrm{D},2}(\vr)
    \end{pmatrix}\,.
\end{align}

A schematic of the effect of the two added terms on the band structure is shown in \prettyref{exfig:theory_figure_interactions}d. We further mimic a cascade 
by shifting two copies of the flat-band subsystem below the Dirac points 
of the band structure and closing the `Fock' gap in the cascaded bands. 

 We present the result of this analysis in \prettyref{exfig:theory_figure_interactions}i, 
 where the filling factor of the flat-band subsystem versus $\Delta_H$ and $\Delta_F$ is 
 plotted for a fixed total filling $\nu = +5$ (corresponding to the edge of 
 the superconducting dome in TPG). We find that in order for scenario ($i$) to apply, i.e. 
 $\nu_\mathrm{flat}\approx +3$ at $\nu=+5$ in TPG, the corresponding parameters are unrealistic. Especially, a 
 $\Delta_F\approx 20$~meV would yield an insulating gap of $40$~meV, 
 which far exceeds a typical correlated insulating gap of few meV experimentally 
 observed in the context of TBG 
 \cite{choiElectronicCorrelationsTwisted2019,kerelskyMaximizedElectronInteractions2019,xieSpectroscopicSignaturesManybody2019,jiangChargeOrderBroken2019}. Notably, for a $\Delta_F$ below 
 $\sim 5$ meV, there is negligible effect of the Fock gap on the flat-band 
 filling redistribution, consistent with our earlier Hartree-only approximation.

\label{methods:interlayer_chem_potential}
\theorysubsubsection{Interlayer screening effects} In the above analysis of 
Hartree-induced band shifting and deformations, we only focused on the effect of the 
in-plane Hartree potential, assuming a homogeneous charge distribution across the layers. 
It is known, however, that in multilayer (untwisted) graphene, interlayer screening 
due to inhomogeneous charge distribution over different layers can be 
important\cite{guineaChargeDistributionScreening2007,koshinoInterlayerScreeningEffect2010}. 
To check the importance of this effect, 
we follow the self-consistent treatment in Ref.~\citenum{koshinoInterlayerScreeningEffect2010} 
and determine the screening-induced potentials on each layer. Particularly, in the language 
of the Hamiltonian in Eq.~\eqref{eq:total_hamiltonian}, we show that even at
zero external displacement field, the induced potentials are not uniform on all
layers and can lead to hybridization between the flat TBG-like and the MLG-like subsystems 
as presented above in section~\ref{methods:mirror_symmetry_chem_variation}.
 
We model the graphene layers as parallel plates with zero thickness and
respective electron charge densities $en_i$. The \emph{local} displacement field between the
$i$th and $(i+1)$th layer is then given by $F_{i,i+1} = e(\sum_{j=1}^{i} n_j - \sum_{j=i+1}^{N} n_j)/(2\epsilon)$, where
$N$ is the total number of layers, and we take $\epsilon=11.15$, the same as the
in-plane dielectric constant determined above. Here, we use the in-plane dielectric 
constant value because electron densities are delocalized over multiple layers and 
cannot be simply treated as some classical charge on a particular layer. Thus, the 
effective dielectric constant should be much larger than the vacuum’s value. 
The above local displacement field produces the local potential
difference between the $i$th and the $(i+1)$th layer $V_{i+1} - V_{i} = -edF_{i,i+1} =-e^2 d/(\sqrt{3}\epsilon L_M^2)\times (\sum_{j=1}^{i} \nu_j - \sum_{j=i+1}^{N} \nu_j)$, where $\nu_j$ is 
the filling fraction projected to the $j$th layer, $d$ is the interlayer distance, and $L_M$ is the moir\'e 
lattice constant.

The role of this mechanism is shown in \prettyref{exfig:theory_figure_charge_inhomo}. 
We find that the self-consistently generated potential differences shift the flat bands 
upwards in energy. Similar to the in-plane Hartree correction, it enables further 
charge filling of the dispersive TBG-like bands, which is in line with scenario ($ii$) 
for the extended TPG superconducting pocket. 
We stress that unlike the in-plane Hartree correction, the out-of-plane Hartree term 
leads to the hybridization of the different subsystems. This hybridization, 
in addition to other effects described in the previous section, may facilitate 
symmetry breaking  or a breakdown of an approximate assignment of flat 
and dispersive TBG-like bands (see the discussion above concerning band mixing, 
section~\ref{methods:mirror_symmetry_chem_variation}), in line with the 
condition for scenario ($iii$).

\section{Possible Origins of the Extended Superconducting Pocket in TPG}
\label{methods:scenario}

Here we present several scenarios that can result in the superconductivity of 
TPG extending to $\nu \approx +5$, and discuss these scenarios in the context of 
experimental observations. We note that in the discussion below, $\nu$ 
denotes the total number of electrons per moir\'{e} site, 
and $\nu_\mathrm{flat}$ denotes the number of electrons per moir\'{e} site 
added to the flat TBG-like bands.

\subsection{Scenario $(i)$: flat TBG-like bands are filled to $\nu_\mathrm{flat} = +3$ at $\nu = +5$}
For TBG and TTG, the strongest superconducting pockets normally start from
$|\nu| = 2$ and end around $|\nu| = 3$. Therefore, a conventional scenario
would suggest that TPG could behave in a similar way, i.e., 
flat TBG-like bands are filled to $\nu_\mathrm{flat} = +3$ when superconductivity 
is diminished at $\nu = +5$. This scenario implies that the additional two electrons per
moir\'e site are distributed in the dispersive TBG- and MLG-like bands due to the
interaction effects discussed in section \ref{methods:Hartree}, with a large portion of the 
charge carriers being hosted by the dispersive TBG-like bands. Since vHs of the dispersive 
TBG-like bands are normally found around half filling, the corresponding 
Hall density signatures are expected to occur at the same filling, 
i.e., $\nu = +5$ in this scenario. However, in the experiment we observe 
vHs signatures originating from the dispersive TBG-like bands 
near $\nu \approx +6$ instead (see \prettyref{exfig:vHS_dispersiveband}). 
This line of reasoning allows us to completely rule out 
scenario $(i)$, therefore, we conclude that superconductivity exceeds 
flat-band filling $\nu_\mathrm{flat} = +3$ for electron-doped TPG. 

\subsection{Scenario $(ii)$: flat TBG-like bands are filled close to 
$\nu_\mathrm{flat} = +4$ at $\nu = +5$} As a result of interactions, 
a fraction of electrons are preferentially distributed in the dispersive TBG- and 
MLG- like bands. It is therefore possible that for total filling 
of $\nu \approx +5$, the flat TBG-like bands are filled close to 
$\nu_\mathrm{flat} \approx +4$, with the extra one electron per moir\'e site 
being distributed in the other bands. We explored this possibility in more detail 
in section \ref{methods:Hartree}, which shows the filling correspondence between
$\nu_\mathrm{flat}$ and $\nu$ for various interaction terms and dielectric constants
(see \prettyref{exfig:theory_figure_interactions}). In this scenario, the modeling
suggests that the filling of the flat bands is nearly four 
($\nu_\mathrm{flat}>+3.8$), which is well outside typical TBG behaviour. 

\subsection{Scenario $(iii)$: flat TBG-like bands are fully filled 
to $\nu_\mathrm{flat} = +4$ before $\nu = +5$ or hybridization of 
different bands obscures the distinction between them} The last scenario 
suggests either that the flat TBG-like bands are fully filled
\emph{before} the suppression of superconductivity, in which case 
superconductivity would exist in the more dispersive bands, \emph{or} that
the distinction between the different TBG- and MLG-like bands 
breaks down due to hybridization (i.e. mixing), even at $D = 0$. As 
discussed in previous sections, such mixing between flat, dispersive 
TBG- and MLG-like bands can happen when mirror symmetry is broken. Moreover, 
layer-to-layer charge inhomogeneity (see \prettyref{exfig:theory_figure_charge_inhomo} 
and section \ref{methods:interlayer_chem_potential})  or distant-layer 
coupling (see section \ref{methods:band_mixing}) allow for 
band hybridization even in the presence of mirror symmetry.

\subsection{Experimental signatures in electron-doped TPG}

\begin{figure}[p]
    \includegraphics[width=16cm]{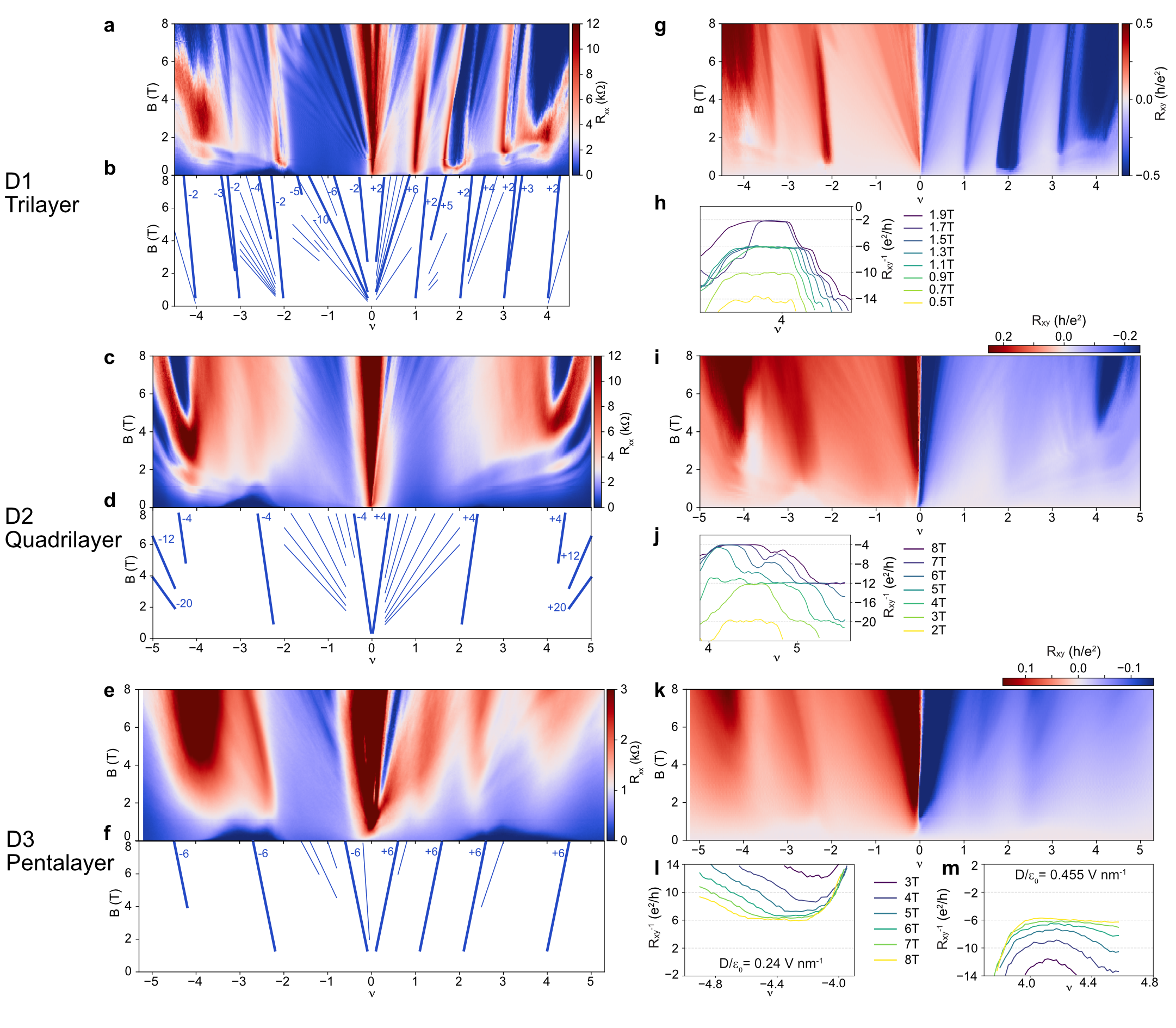}
    \centering
    \caption{{\bf Fan diagrams at zero $D$ field and the Hall conductance quantization around
        $|\nu| = 4$.} {\bf a}--{\bf f}, $R_{xx}$ measured as a function of $B$
      field and $\nu$ from trilayer to pentalayer ({\bf a}, {\bf c}, {\bf e}). The main sequences of the fan
      diagrams are labelled at the bottom of $R_{xx}$ ({\bf b}, {\bf d}, {\bf f}). Landau levels from the dispersive bands are visible as $R_{xx}$ oscillations at low $B$ fields in the fan diagrams. {\bf g}--{\bf m}, $R_{xy}$ measured as a function of $B$ field and $\nu$ from
      trilayer to pentalayer ({\bf g}, {\bf i}, {\bf k}). Below these plots, we show Hall 
      conductance around $|\nu| = 4$ ({\bf h}, {\bf j}, {\bf l}, {\bf m}). The 
      layer number $n$ determines the resulting quantization. Since the dispersive bands of 
      $n$-layer twisted graphene consist of $n-2$ Dirac-like cones (at low energies), 
      the $|\nu| = 4$ quantization is therefore expected to follow monolayer 
      graphene sequence ($\pm2$, $\pm6$, $\pm10$,..., $\times e^{2}/h$) multiplied 
      by $n-2$. The plateaus in TTG and TQG clearly show this trend, while in TPG 
      only the first plateau is observed. These observations however confirm the 
      number of layers in each sample.}
\label{exfig:Fan diagram}
\end{figure}

\begin{figure}[p]
    \includegraphics[width=15cm]{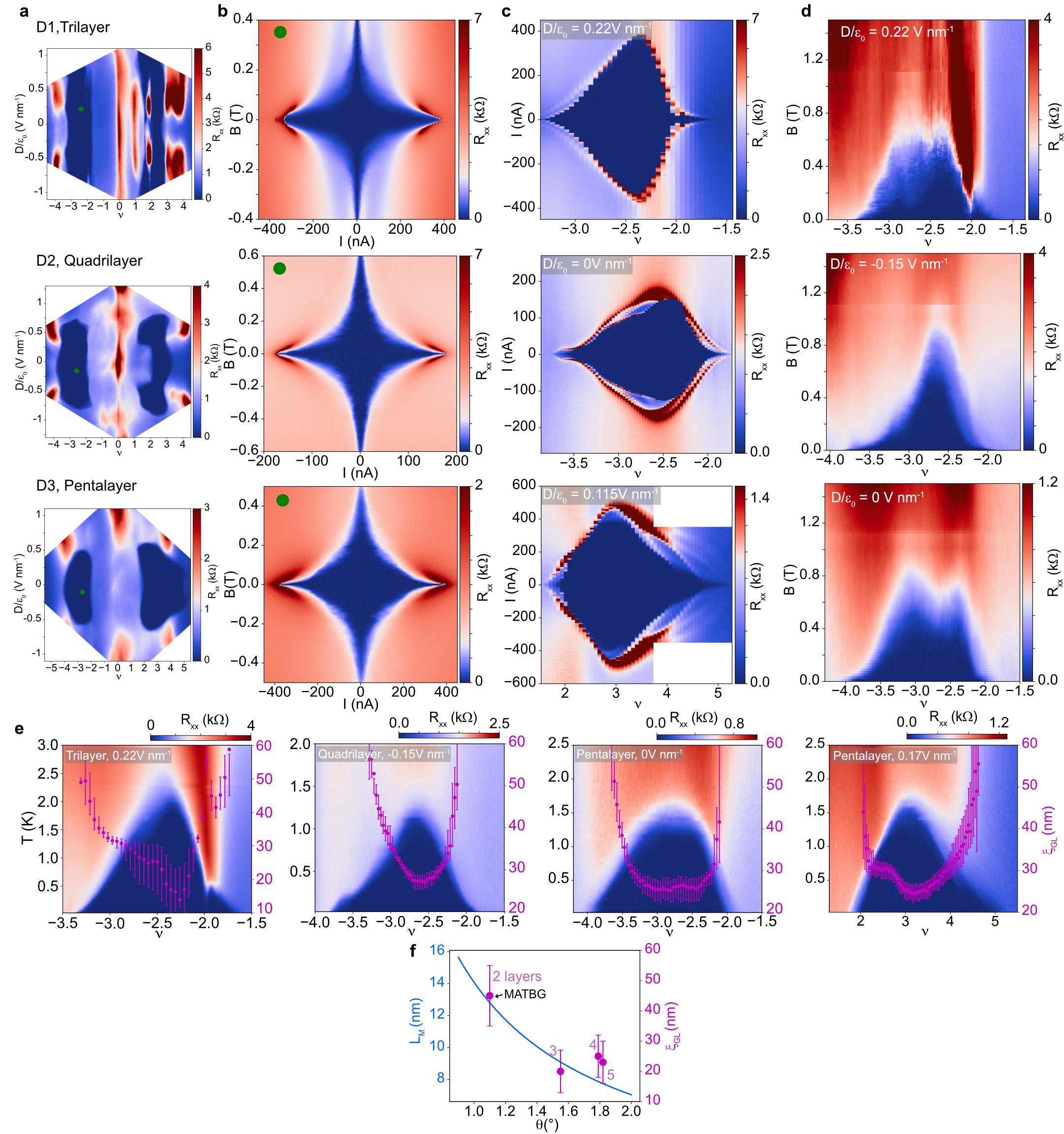}
    \centering
    \caption{{\bf Fraunhofer patterns, $I_c$, $B_c$ and coherence lengths of TTG, TQG and TPG.} 
    Column {\bf a} shows $R_{xx}$ versus $\nu$ and $D$ phase diagrams, and the green dots indicate the positions where the corresponding 
    Fraunhofer interference patterns ({\bf b}) are measured for D1--D3.
    Column {\bf c} shows the critical current $I_{c}$ versus $\nu$ at the
    optimal $D$ fields for D1--D3. Column {\bf d} shows $R_{xx}$ versus $\nu$ and $B$ around $\nu=-2$ 
    for D1--D3, highlighting the high critical magnetic fields in these systems.
      Superconductivity in the twisted graphene multilayers has a higher $B_{c}$
      ($\sim0.8$~T or higher) than in TBG. {\bf e}, Ginzburg--Landau coherence
      lengths $\xi_\mathrm{GL}$ versus $\nu$ for all three devices around
      $|\nu| = 2$, superimposed on the $R_{xx}$ versus $T$ and $\nu$ plots. 
      {\bf f}, $\xi_\mathrm{GL}$ and moir\'e wavelength $L_\mathrm{M}$ versus 
      twist angle of different layers, suggesting a possible 
      relation between the two length scales.}
\label{exfig:Fraunhofer}
\end{figure}

\begin{figure}[p]
    \includegraphics[width=16cm]{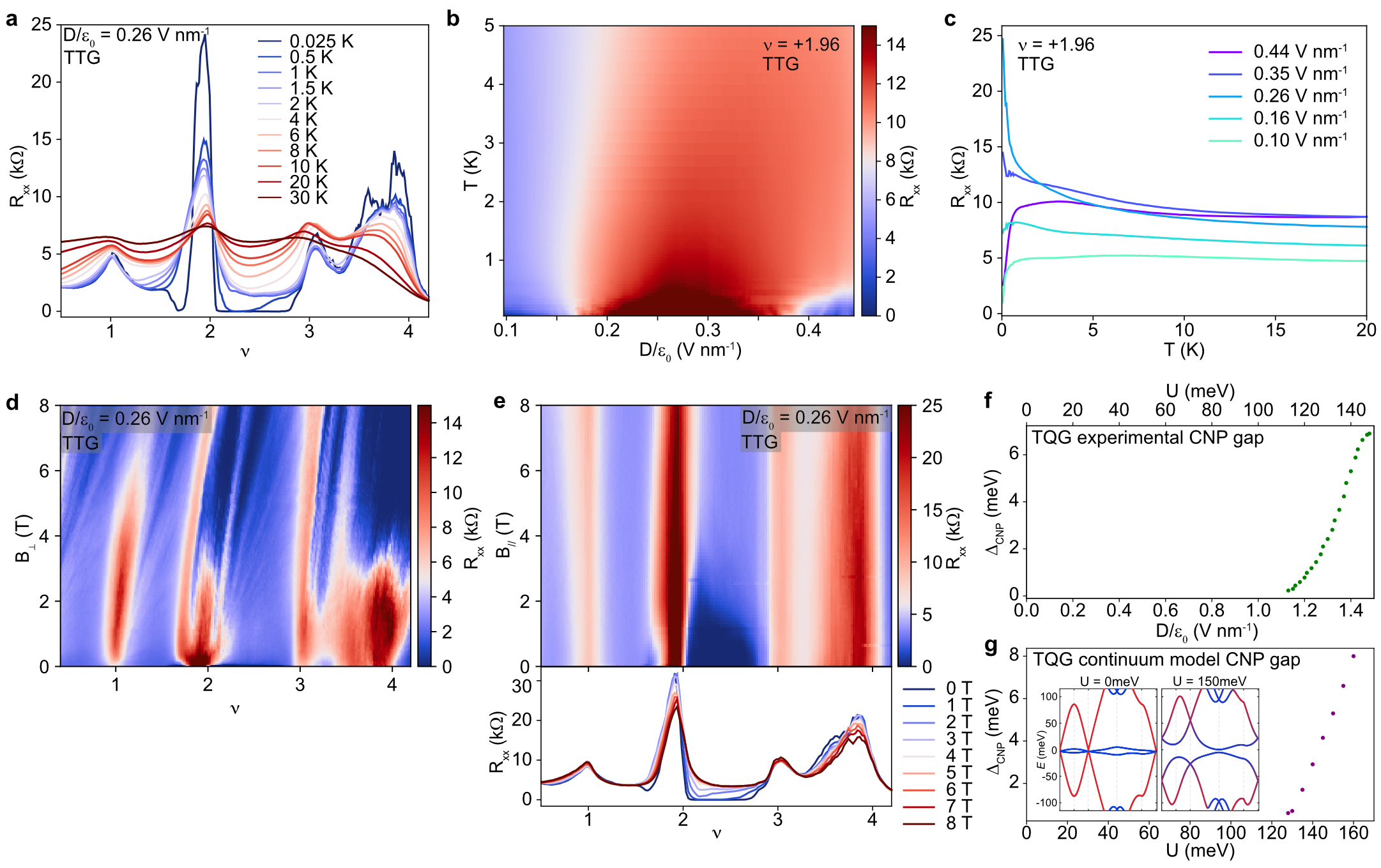}
    \centering
    \caption{{\bf Insulating behaviour in TTG and TQG.} {\bf a}, Line cuts of
      $R_{xx}$ versus $\nu$ for a range of temperatures at $D/\epsilon_0 = 0.26~\text{V
        nm}^{-1}$ on the electron side for TTG. {\bf b}, $R_{xx}$
      versus $D$ and temperature at $\nu = +1.96$ in TTG. {\bf c}, Line cuts at
      different $D$ fields from {\bf b}. Out-of-plane ({\bf d}) and
      in-plane ({\bf e}) $B$ field dependence of $R_{xx}$ versus $\nu$ at
      $D/\epsilon_0 = 0.26~\text{V nm}^{-1}$ in TTG. The $\nu = +2$ correlated
      insulator is suppressed by both in-plane and out-of-plane $B$ field. {\bf
        f}, Experimental charge-neutrality gap of TQG as a function of $D$ field,
      and {\bf g}, the continuum-model gap as a function of potential difference $U$. Inset, single-particle band structure of TQG (slightly above the magic angle) at $U = 0~\text{meV}$ and $150~\text{meV}$, respectively. We see a good
      match between experiment and theory when converting $D$ into $U$ with an empirical
      factor: $U = 0.1\times (n-1)\times0.33~{\rm nm}\times eD$, where $n-1$ is
      the number of graphene interfaces. }
\label{exfig:CI_TTG}
\end{figure}

\begin{figure}[p]
    \includegraphics[width=16cm]{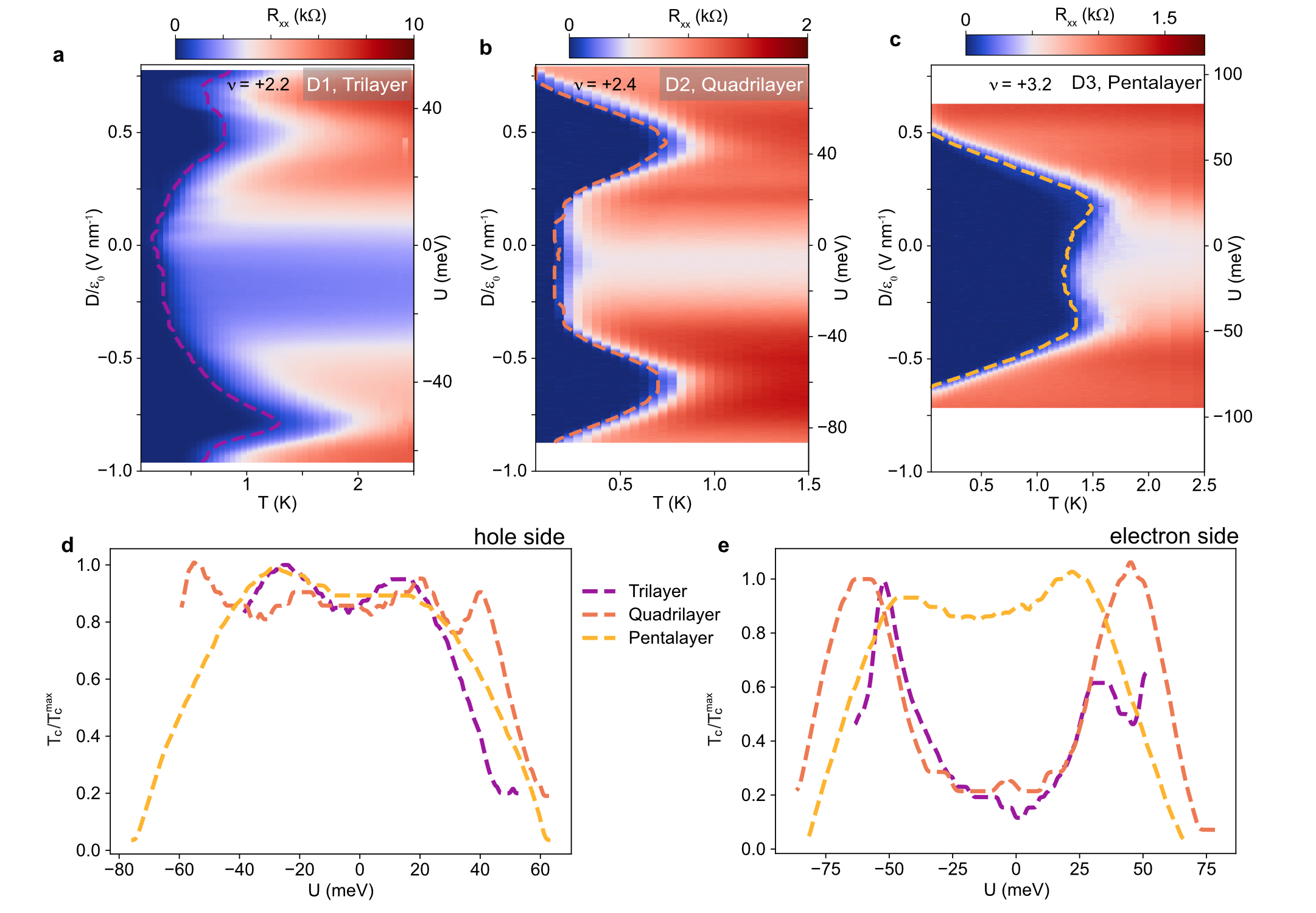}
    \centering
    \caption{{\bf Evolution of superconducting critical temperature $T_c$ with $D$ field 
    around optimal doping.} {\bf a}--{\bf c}, $R_{xx}$ as a function of $T$ and $D$ field for
      D1--D3 at filling factor $\nu = +2.2$, $+2.4$, and $+3.2$, respectively.
      Superconducting $T_c$ is indicated by a dashed line that delineates $10\%$
      of the normal state resistance (see section
      \ref{methods:T_B_dependence} for details). {\bf d},{\bf e},
      $T_{c}/T_{c}^{max}$ versus potential energy difference $U$ for TTG, TQG,
      and TPG around hole-side ({\bf d}) and electron-side ({\bf e}) optimal doping,
      respectively. $U$ is converted from $D$ using
      $U = 0.1\times (n-1)\times0.33~{\rm nm}\times eD$, where $e$ is the
      electron charge and $n-1$ is the number of graphene interfaces.}
\label{exfig:Tc_vs_D_electron}
\end{figure}

\begin{figure}[p]
    \includegraphics[width=16cm]{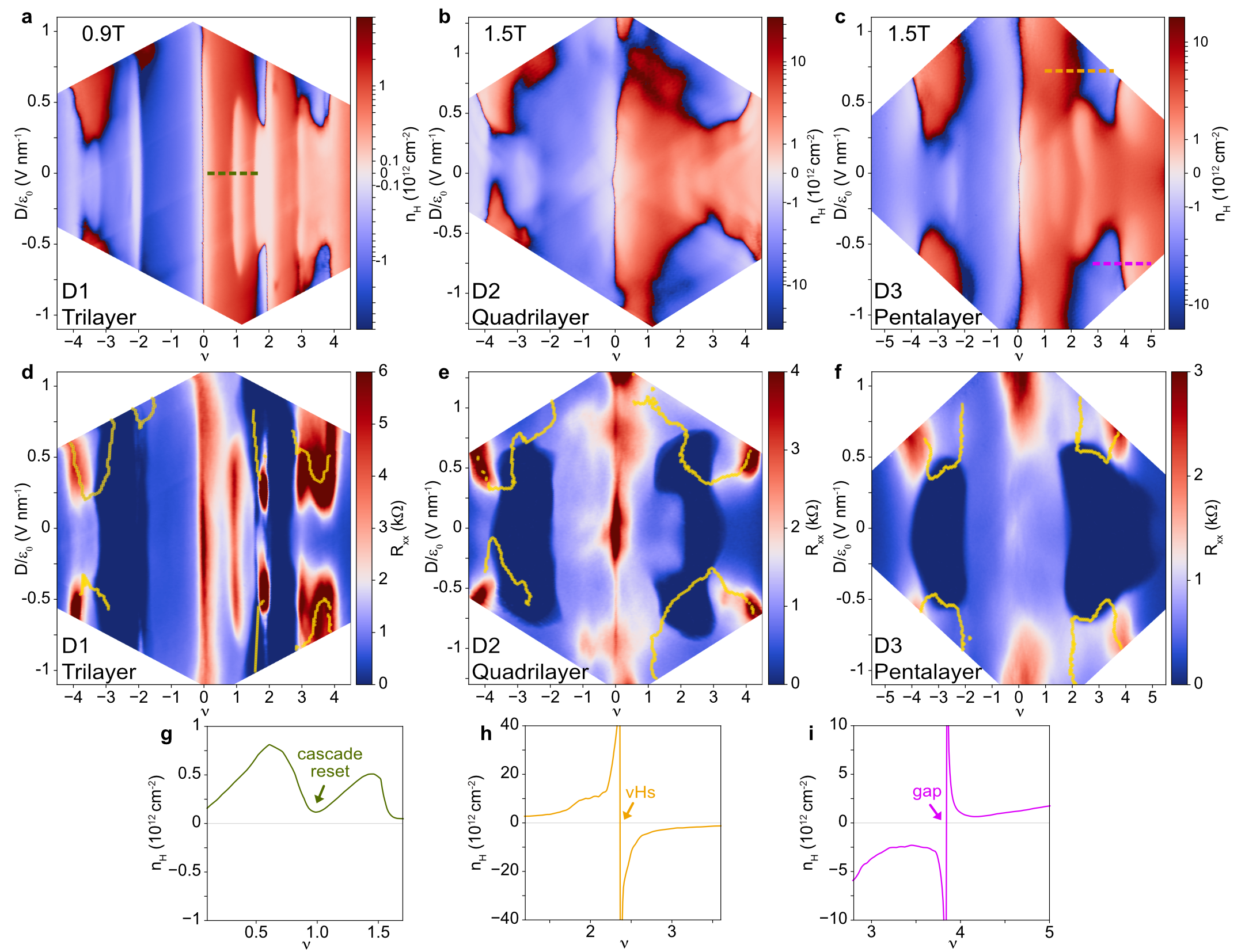}
    \centering
    \caption{{\bf Hall density $\nu$--$D$ maps and the positions of vHs/`gap’ features.} 
    {\bf a}--{\bf f}, Hall density ({\bf a}--{\bf c}) and $R_{xx}$ ({\bf d}--{\bf f}) 
    as a function of $\nu$ and $D$ for TTG, TQG, and TPG. Hall density maps are measured at $B=0.9$~T, $1.5$~T, and
      $1.5$~T, respectively. Yellow lines in {\bf d}--{\bf f} track the evolution of
      vHs/`gap’ features where Hall density changes sign. {\bf g}--{\bf i}, Examples of Hall density 
      near the cascade transition reset ({\bf g}), the vHs ({\bf h}), and the `gap’ 
      ({\bf i}) following the definitions in Ref.~\citenum{parkTunableStronglyCoupled2021}. 
      Filling ranges for the line cuts are marked by the corresponding colored dashed 
      lines in {\bf a} and {\bf c}.}
\label{exfig:Hall density and vHs}
\end{figure}

\begin{figure}[p]
    \includegraphics[width=16cm]{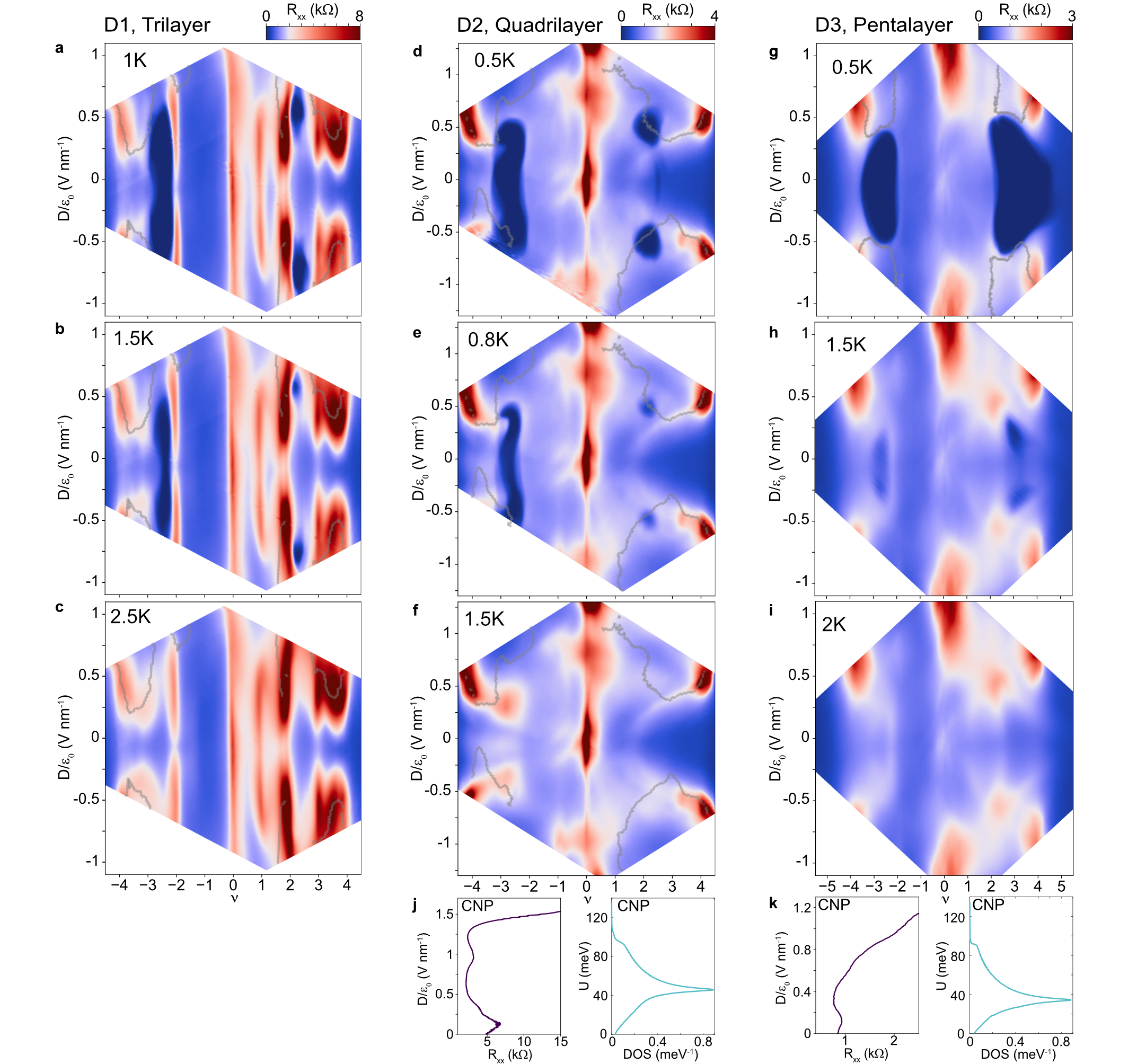}
    \centering
    \caption{{\bf $R_{xx}$ as a function of $\nu$ and $D$ at different temperatures.} 
    {\bf a}--{\bf i}, $R_{xx}$ as a function of $\nu$ and $D$ measured at different
    temperatures for TTG ({\bf a}--{\bf c}), TQG ({\bf d}--{\bf f}), and 
    TPG ({\bf g}--{\bf i}). Grey lines track the 
    evolution of the vHs/`gap’ features. {\bf j}, The plot on the left shows line cut of $R_{xx}$ 
    versus $D$ at charge neutrality for TQG. The plot on the right shows corresponding density of states (DOS) at charge-neutrality point (CNP) calculated using non-interacting continuum model. In the regions where DOS is high, resistance is expected to be low and vice versa. {\bf k}, 
    Equivalent plots as in {\bf j} for TPG. }
\label{exfig:gatemapT}
\end{figure}

Experimentally, starting from low $D$ fields, we observe a drop 
in Hall density at $\nu\approx+4$ which surprisingly does not affect 
superconductivity in any abrupt way (superconductivity continuously evolves and 
is present until $\nu\approx+5$). As the $D$ field is increased, this Hall density drop 
is gradually replaced by a transition where Hall density changes sign (\prettyref{exfig:Hall_density_TPG}). The 
high $D$-field transition can be interpreted as a `gap' feature emerging 
in the band structure similar to TTG\cite{parkTunableStronglyCoupled2021}.
Further measurements of $R_{xx}$ show that the corresponding $\nu\approx+4$
feature does not shift with temperature (\prettyref{fig:Fig4}a) and is significantly broadened 
at high $B$ fields, resembling the feature associated with the flat-band gap in TTG (\prettyref{exfig:Fan diagram}a and e). 
These observations indicate that the $\nu=+4$ feature is naturally explained as either marking the end of the flat bands or resulting from band details due to hybridization, which is in line with the 
scenario $(iii)$. In this context, the alternative possibility that $\nu=+4$ corresponds to 
a flavor-polarization reset at $\nu_\mathrm{flat}=+3$ is highly unlikely. Finally, we note that 
this line of argument cannot fully rule out scenario $(ii)$ due to the potential presence of 
small dispersive pockets in the flat bands that may remain unfilled near $\nu=+4$.

\clearpage
\beginsupplement

\clearpage

\clearpage

\clearpage

\clearpage

\clearpage

\clearpage

\clearpage

\begin{figure}[p]
    \includegraphics[width=16cm]{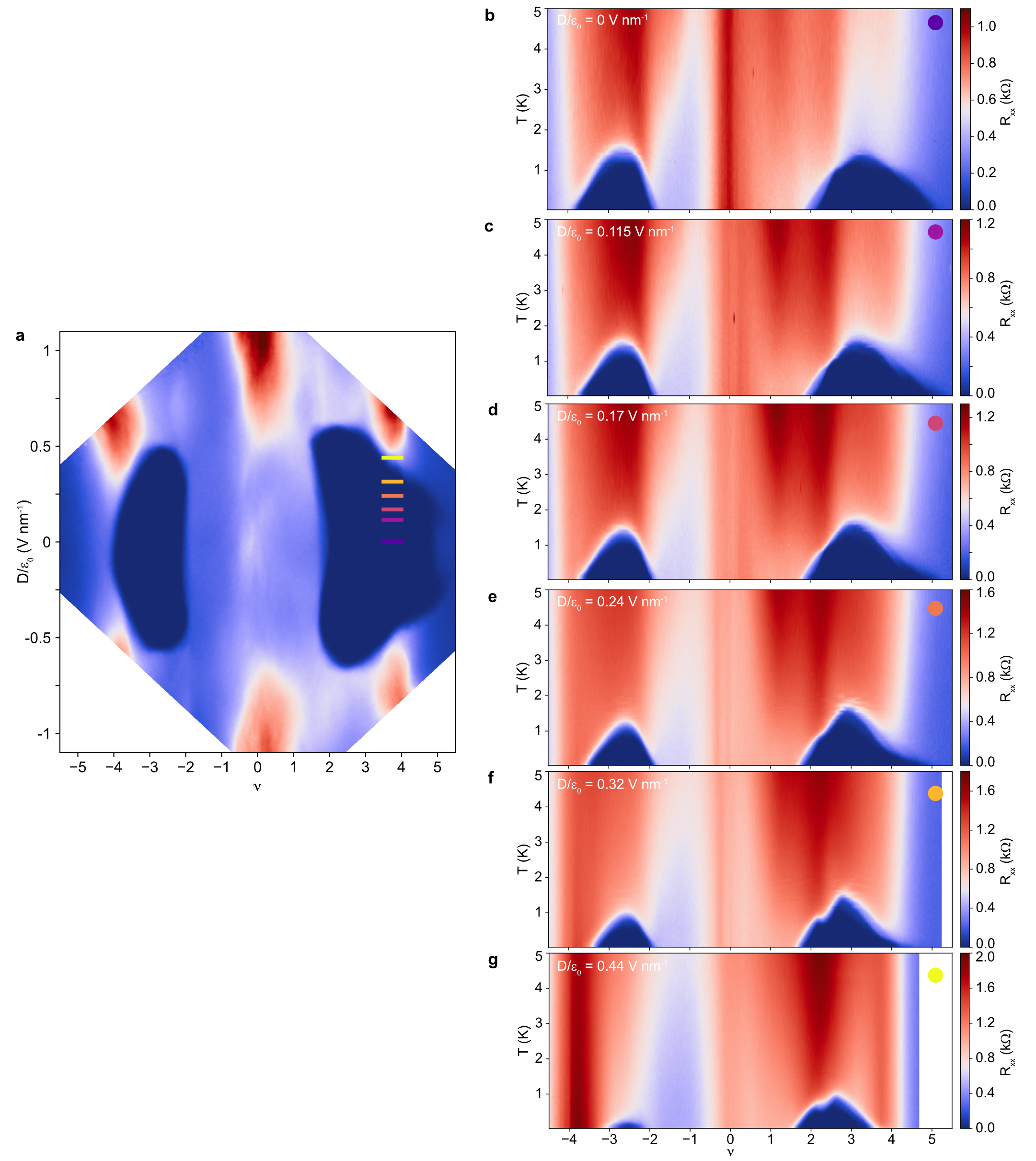}
    \centering
    \caption{{\bf Evolution of superconducting $\nu$--$T$ domes with displacement field $D$ in TPG.} 
    {\bf a}, $R_{xx}$ as a function of $\nu$ and $D$ in TPG. 
    {\bf b}--{\bf g}, $R_{xx}$ versus $\nu$ and temperature at 
    different $D$ fields, and $D$
    fields are marked with colored bars in {\bf a}. }
\label{exfig:SC_Ddepend}
\end{figure}
\clearpage

\begin{figure}[p]
    \includegraphics[width=16cm]{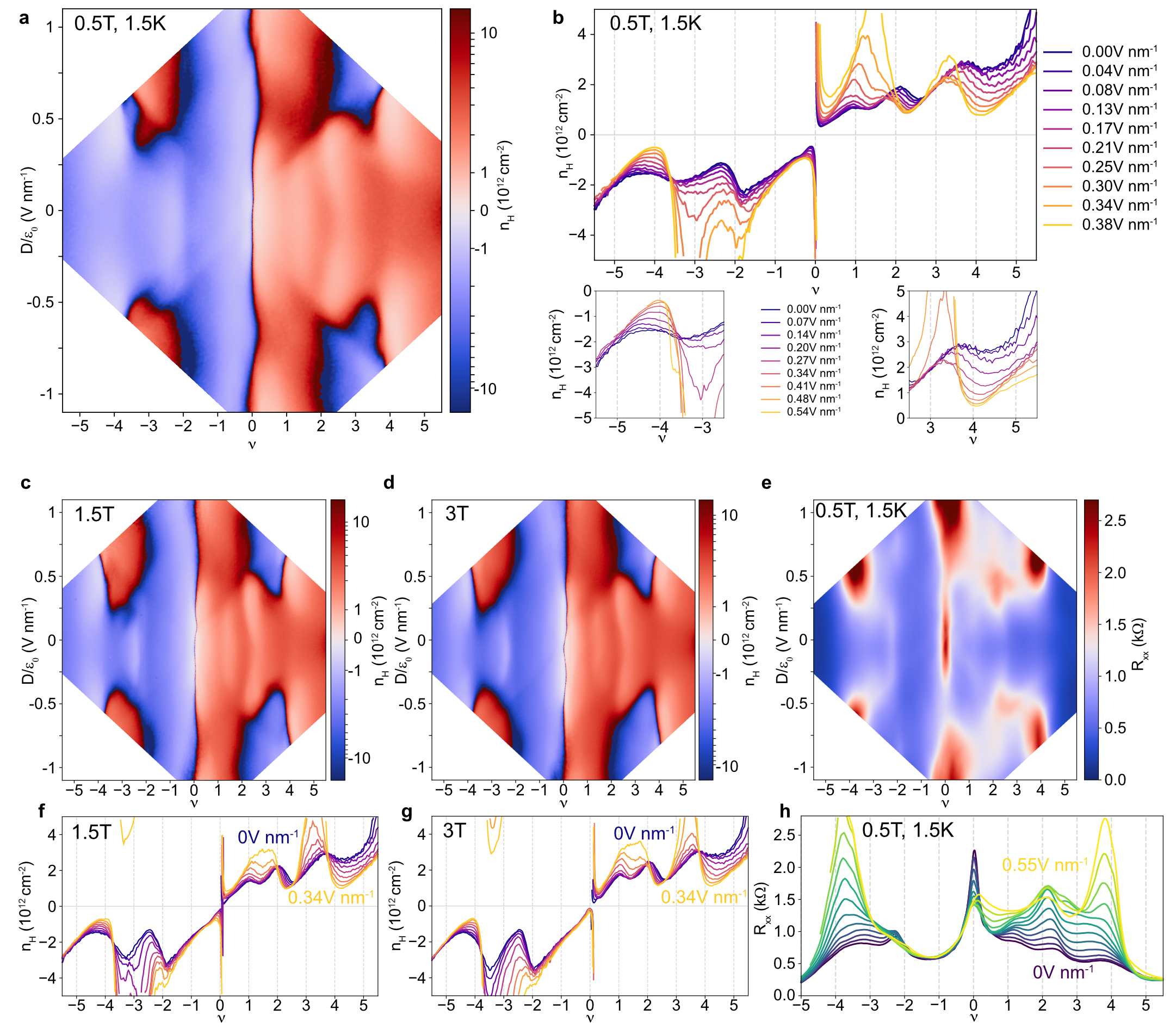}
    \centering
    \caption{{\bf Hall density and $R_{xx}$ as a function of $\nu$ and $D$ measured 
    at different $B$ fields in TPG.} {\bf a}, Hall density versus $D$ and $\nu$ at $B=0.5$~T.
    {\bf b}, Line cuts from {\bf a}. Panels below zoom in on the evolution of Hall density 
    resets near $|\nu|=4$. {\bf c},{\bf d}, Hall density versus
    $D$ and $\nu$ measured at $B=1.5$~T ({\bf c}) and $3$~T ({\bf d}), with respective line cuts shown in {\bf f} and {\bf g}. {\bf e}, $R_{xx}$ versus $D$ and $\nu$ measured at 
    $T=1.5$~K, $B=0.5$~T (line cuts are shown in {\bf h}). From all the above 
    line cuts, Hall density resets and $R_{xx}$ resistive features consistently 
    exist around $\nu = +4$. }
\label{exfig:Hall_density_TPG}
\end{figure}
\clearpage

\begin{figure}[p]
    \includegraphics[width=16cm]{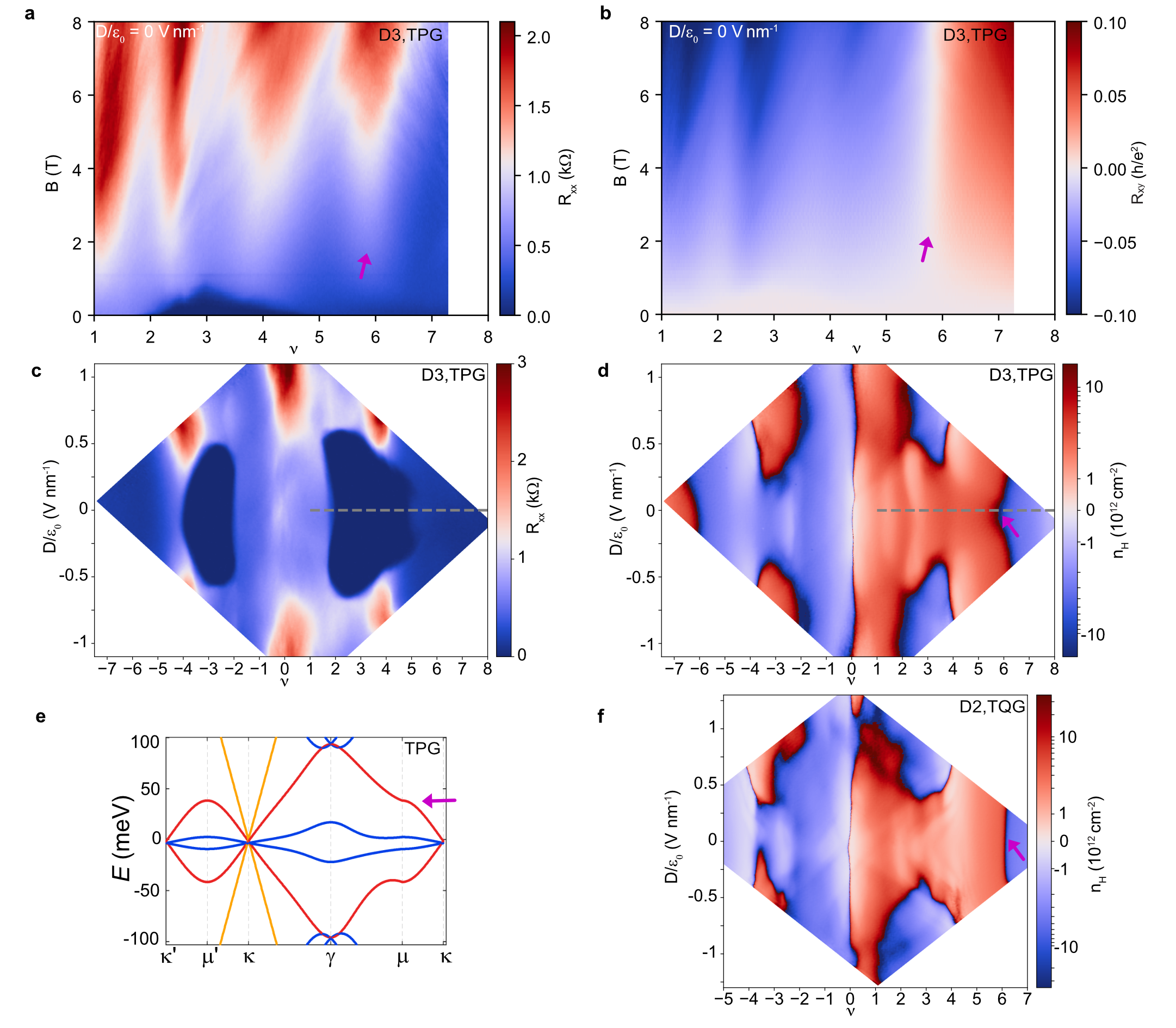}
    \centering
    \caption{{\bf Plots in a broader filling factor range and vHs of dispersive 
    TBG-like bands in TPG and TQG.} {\bf a},{\bf b}, $R_{xx}$ and $R_{xy}$ as 
    a function of $\nu$ and $B$ field measured at zero $D$ field for TPG. 
    The sign change in $R_{xy}$ around $\nu = +6$ (marked by arrows in {\bf a} and {\bf b})
    indicates vHs. {\bf c},{\bf d}, $R_{xx}$ ({\bf c}) and Hall density ({\bf d}) as a function of $D$ and $\nu$ with gray dashed lines indicating $\nu$ linecuts (at $D=0$) where plots in {\bf a} and {\bf b}
    are taken. {\bf e}, Band structure of TPG calculated using non-interacting model. Arrow
    indicates the position where vHs from dispersive TBG-like bands is expected. {\bf f}, 
    Hall density as a function of $\nu$ and $D$ for TQG. As in TPG, Hall density changes sign 
    near $\nu=+6$ indicating the vHs from dispersive 
    TBG-like bands in TQG.}
\label{exfig:vHS_dispersiveband}
\end{figure}
\clearpage

\begin{figure}[p]
    \includegraphics[width=16cm]{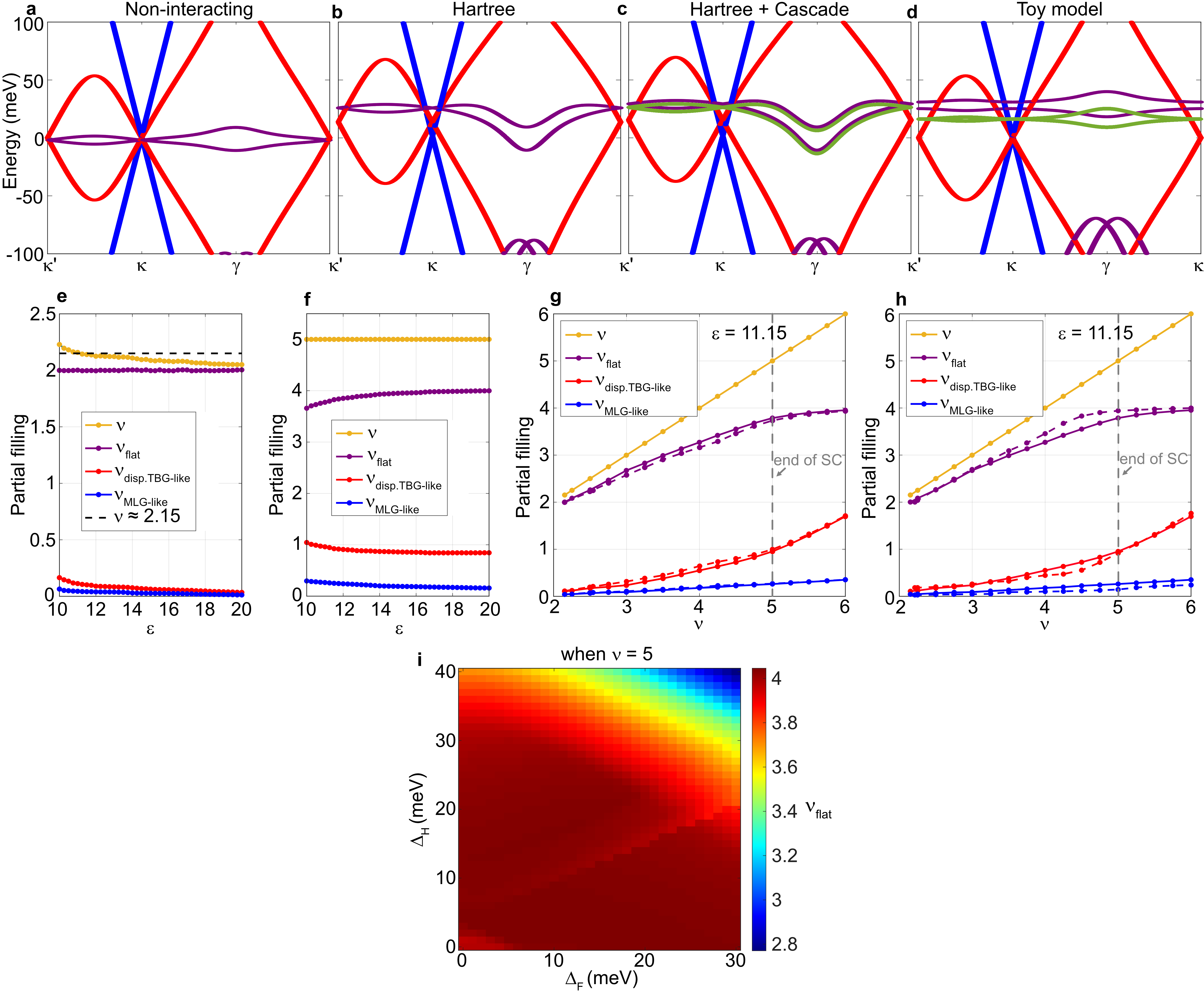}
    \centering
    \caption{{\bf The role of interactions in TPG.} {\bf a}-{\bf
        d}, Depiction of different approximation schemes used to 
        understand the role of interactions in TPG. Note that the 
        Hartree correction shifts the flat band (purple) up in energy. 
        Cascaded bands in {\bf c} and {\bf d} are shown in green. {\bf d} 
        corresponds to a simple toy model of Hartree and Fock effects 
        characterized by a Hartree shift ($\Delta_H$) and a Fock gap 
        ($\Delta_F$) (see section \ref{methods:hartree_and_fock_corrections}). 
        {\bf e},{\bf f}, Partial filling of each subsystem versus dielectric 
        constant $\epsilon$ for a fixed flat-band filling $\nu_\mathrm{flat}=+2$ 
        ({\bf e}) and a fixed total filling $\nu = +5$ ({\bf f}), respectively. 
        {\bf g}, Partial filling of each subsystem versus total filling $\nu$ 
        for a fixed dielectric constant $\epsilon = 11.15$. Here, solid (dashed) 
        lines correspond to a cascaded (uncascaded) solution with the cascade 
        solution enabling higher filling of the flat-band subsystem as 
        discussed in the text. {\bf h}, Similar to {\bf g} but the solid 
        (dashed) lines correspond to a solution at potential 
        difference $U = 0$~meV ($U = 34$~meV). {\bf i}, Filling of the 
        flat-band subsystem as a function of $\Delta_H$ and $\Delta_F$ 
        at a fixed total filling $\nu=+5$ 
        (see section \ref{methods:hartree_and_fock_corrections}).}
\label{exfig:theory_figure_interactions}
\end{figure}
\clearpage

\begin{figure}[p]
    \includegraphics[width=16cm]{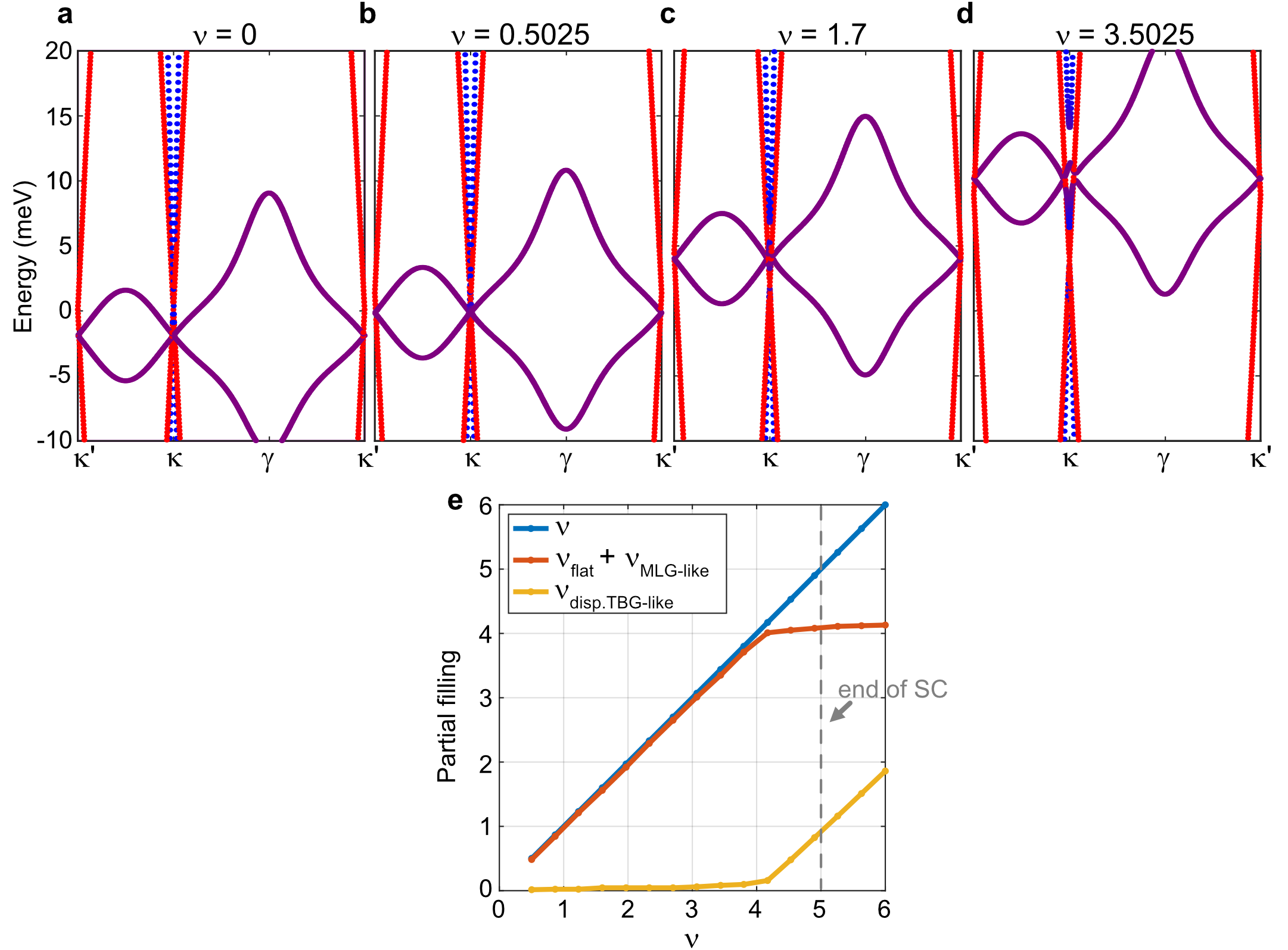}
    \centering
    \caption{{\bf The role of interlayer inhomogeneous charge distribution in TPG.} 
    {\bf a}--{\bf d}, Band structure of TPG at different filling factors with 
    an interlayer inhomogeneous charge distribution. Note that as filling 
    is increased, the flat band is slightly shifted and also hybridizes 
    with the MLG-like band. {\bf e}, Partial filling of different subsystems 
    as a function of total filling $\nu$ with the effect of interlayer 
    inhomogeneous charge distribution. Note a small charge redistribution 
    between $+2 \lesssim \nu \lesssim +5$. Here, flat TBG-like and MLG-like 
    subsystems are plotted together to demonstrate the emergent hybridization.}
\label{exfig:theory_figure_charge_inhomo}
\end{figure}
\clearpage

\end{document}